\documentclass[%
 reprint,
superscriptaddress,
 amsmath,amssymb,
 aps, prresearch,
 longbibliography
]{revtex4-2}

\usepackage{comment}
\usepackage{mathtools}
\usepackage{graphicx}
\usepackage{dcolumn}
\usepackage{float}
\usepackage{bm}
\usepackage{amsmath}
\usepackage{amsthm}

\usepackage[colorlinks]{hyperref} 
\begin{document}

\preprint{APS/123-QED}

\title{Residual dynamics determine dissipation in strongly coupled linear Langevin systems}


\author{Vansh Kharbanda}
\author{Anton F. Burnet}
\author{Benedikt Sabass}%

\affiliation{Department of Physics, Technical University of Dortmund, 44227 Dortmund, Germany}

\affiliation{
 Faculty of Physics and Center for NanoScience, Ludwig-Maximilians-Universit\"at M\"unchen, 80752 Munich, Germany
}%
\affiliation{
 Department of Veterinary Sciences, Ludwig-Maximilians-Universit\"at M\"unchen, 80752 Munich, Germany
}




\date{\today}

\begin{abstract}
Interactions shape the properties of many physical and biological nonequilibrium systems, but whether collective operation raises or lowers entropy production remains difficult to predict. Here, we show that strong coupling suppresses irreversible circulation associated with the directly coupled degrees of freedom. In identical two-dimensional systems, this suppression is monotonic for nonzero coupling modes and guarantees a monotonic decrease of the entropy production rate (EPR), with the corresponding probability currents vanishing in the strong-coupling limit. In higher dimensions, suppressing the circulation associated with the coupled degrees of freedom does not generally eliminate all irreversible circulation, since the remaining degrees of freedom can still sustain nonequilibrium currents. The EPR of these residual dynamics sets the strong-coupling limit, providing a criterion for whether a given choice of coupled variables leads to lower dissipation than the uncoupled system. The approach to this limit depends on how the finite-coupling current correction aligns with the residual current. Further, we show that convergence to residual dynamics in the strong-coupling limit also applies to coupled non-identical systems. Finally, we demonstrate the predicted EPR reduction in nonlinear models of vertebrate hair-cell bundles and bacterial chemotaxis.

\end{abstract}

\maketitle

\section{Introduction}

Biological systems continually dissipate energy to sustain their functions~\citep{Esposito2019Thermodynamic,Ouldridge2017,Bassett2021,Dabelow2019}. Many operate collectively and far from equilibrium with high functional efficiency~\citep{Hwang2018Energetic,Yang2021,Shimokawa2007Enhancement,Slanina2008}. Across scales, a recurring theme is that collective organization can shape the energetics~\citep{Julicher1995,Shimokawa2007Enhancement,Burnet2026Energetics,zhang2020nonequilibrium,Kharbanda_2024,Slanina2008}. This raises a general question: when can coupling reduce the thermodynamic cost of nonequilibrium operation, and by what mechanism?

A natural measure of this cost is the steady-state entropy production rate (EPR)~\citep{Shiraishi2016,Pietzonka2016Universal,Seifert_IFT}. Beyond its foundational role in nonequilibrium thermodynamics~\citep{Gallavotti_Cohen,Esposito_2010,Seifert_IFT,Jarzynski2011,Crook_1999,Ciliberto_2017}, the EPR also constrains information processing, control and adaptation in stochastic systems~\citep{dunkel2021improved,Seifert_2016_Cost,ilker2022shortcuts,Ito2015,Parrondo2015,Ito_housekeeping,Ito_wasserstein,Lan2012,Sartori2014}. Understanding how coupling changes the EPR is therefore central to both interpreting biological efficacy and designing low-dissipation synthetic and engineered networks.

Recent work has increasingly addressed how interactions reshape dissipation in collective nonequilibrium systems. In interacting oscillator systems, approaches from stochastic thermodynamics have connected synchronization and collective dynamics to thermodynamic performance and energetic costs~\citep{Meibohm2024Minimum,imparato2015stochastic,izumida2016energetics,lee2018thermodynamic,Burnet2026Energetics}. Cooperative effects have also been studied extensively in molecular-motor systems, where interactions between motors can modify collective force generation, transport, and energetic efficiency~\citep{Julicher1995,Julicher1998_motors,Slanina2008,kaya2017coordinated}. For example, interactions between kinesin motors have been shown to enhance their efficiency at maximum power relative to the single-motor case~\citep{Imparato_2012}. In chemosensory systems, cooperative interactions have been shown to influence sensing accuracy, sensitivity, and energetic dissipation~\citep{Kharbanda_2024,Hathcock2024Time-reversal}. Additionally, hydrodynamic coupling has been shown to alter irreversibility in mesoscopic systems~\citep{Das2025Irreversibility}. Together, these studies demonstrate that interactions can modify nonequilibrium dissipation across a broad range of systems. However, existing results are often model-specific and can depend sensitively on the coupling architecture or driving protocol~\citep{Imparato_2012,Kharbanda_2024,kaya2017coordinated}.

Linear Langevin systems constitute a particularly useful class of stochastic dynamics, describing a wide range of physical systems~\citep{Hasselmann1976,vanZon2004} and biological systems~\citep{Mura2018,Gilson2023,Sherman1988}, and arising naturally from the linearization of nonlinear dynamics around stable fixed points~\citep{Kwon2005,Aslyamov2026}. In general, the dynamics of linear Langevin systems are described by Ornstein--Uhlenbeck (OU) processes~\citep{Uhlenbeck1930,Wang1945}, whose steady-state statistics and stochastic thermodynamics are well established when the dynamics are stable~\citep{Landi2013,Godreche2019Characterising,duBuisson2023}. Recent work has continued to uncover nontrivial connections between entropy production, oscillatory dynamics, spectral properties, and nonnormality~\citep{Sekizawa2024,Fyodorov2025,Nagayama2026}. Yet even within this seemingly simple and ubiquitous class, several fundamental questions remain open. 

One such question concerns the conditions under which coupling between these systems suppresses or enhances dissipation. In particular, it is not immediately clear how coupling reorganizes irreversible probability currents and how these changes affect the EPR. The strong-coupling regime offers a useful setting in which to investigate these questions. At sufficiently large coupling strengths, the relaxation rates of the directly coupled degrees of freedom (DoF) can effectively increase, leading to a separation of timescales between the coupled and uncoupled variables. This separation provides a way to examine how coupling affects the irreversible circulation associated with each set of variables and, consequently, their respective contributions to the EPR of the full system.

In this article, we study linear Langevin subsystems coupled through reciprocal linear interactions. We show that strong coupling suppresses irreversible circulation associated with the directly coupled DoF. In two dimensions, this suppression is monotonic for nonzero coupling modes and guarantees a monotonic decrease of the corresponding EPR contributions, while their probability currents vanish in the strong-coupling limit. In higher dimensions, however, the remaining DoF can sustain residual irreversible circulation. The EPR associated with these residual dynamics determines the strong-coupling limit, thereby providing a criterion for whether a strongly coupled system dissipates less than the corresponding uncoupled system. The approach to this limit depends in part on the alignment between the residual irreversible current and the leading finite-coupling correction induced by the coupled DoF. This picture naturally extends to non-identical subsystems, where parameter heterogeneity modifies the residual dynamics and can alter the limiting EPR. We use nonlinear models of bacterial chemotaxis and vertebrate hair-cell bundles to further illustrate this mechanism beyond the linear setting.

\section{Setup}

\begin{figure}
    \centering
    \includegraphics[width=\linewidth]{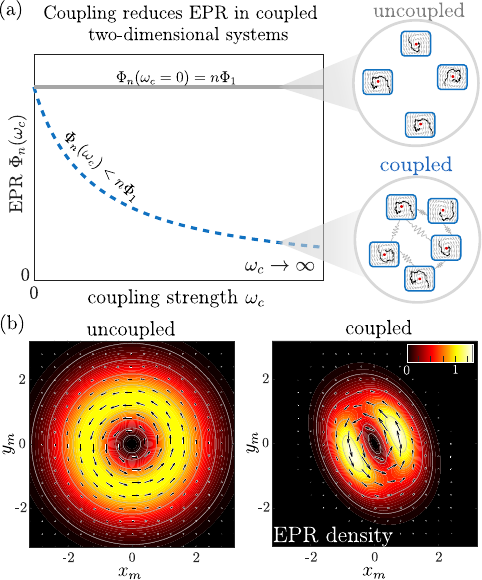}
    \caption{Coupling reduces EPR in two-dimensional systems.
(a) Coupling suppresses irreversible circulation in nonzero
coupling modes, reducing the total EPR from its uncoupled
value $\Phi_n(0)=n\Phi_1$ towards its strong-coupling limit.
(b) Stationary probability current (arrows) overlaid on
the local EPR density,
$\phi_m = \mathbf j_m^\top \mathbf D_0^{-1}
\mathbf j_m/p(\mathbf x_m)$,
for a representative nonzero mode at $\omega_c=0$ and
$\omega_c=3$, respectively. Increasing coupling suppresses irreversible circulation and confines the probability current to a narrower region of phase space, thereby reducing the associated dissipation.
Parameters: $a=1$, $b=3$, $c=3$, $d=1$, and $D_x=D_y=1$.}
    \label{fig:panel1}
\end{figure}
\subsection{Uncoupled dynamics and EPR in linear Langevin systems}

We consider $n$ identical $d$-dimensional overdamped linear Langevin systems, where the uncoupled dynamics of system $i$ are governed by
\begin{equation}
    \boldsymbol{\Gamma}_0 \frac{\mathrm{d}\mathbf{x}_i(t)}{\mathrm{d}t}
    =
    \boldsymbol{\Lambda}_0\mathbf{x}_i(t)
    +
    \sqrt{2\boldsymbol{\Gamma}_0\mathbf T_0^{\mathrm{eff}}}\,
    \boldsymbol{\eta}_i(t).
    \label{eq:langevin}
\end{equation}
Here, $\mathbf{x}_i\in\mathbb R^d$ denotes the degrees of freedom, $\boldsymbol{\eta}_i$ are independent standard Gaussian white-noise vectors, $\boldsymbol{\Gamma}_0$ is a positive-definite diagonal friction matrix, and $\boldsymbol{\Lambda}_0$ is the linear force matrix. The effective-temperature matrix is $\mathbf T_0^{\mathrm{eff}}=\operatorname{diag}(T_1^{\mathrm{eff}},\ldots,T_d^{\mathrm{eff}})$, with $T_l^{\mathrm{eff}}>0$ and $k_{\mathrm B}=1$. We define the drift and diffusion matrices as
 $\mathbf M_0=\boldsymbol{\Gamma}_0^{-1}\boldsymbol{\Lambda}_0,$ and $ \mathbf D_0=\boldsymbol{\Gamma}_0^{-1}\mathbf T_0^{\mathrm{eff}},$
and assume that $\mathbf M_0$ is Hurwitz stable, meaning that all its eigenvalues have negative real parts \citep{Godreche2019Characterising}.

Since the subsystems are independent and identical, the total uncoupled EPR is $\Phi_n(0)=n\Phi_1$, see Fig.~\ref{fig:panel1}(a), where 
\begin{equation}
    \Phi_1
    =
    \int_{\mathbb R^d}
    \frac{\mathbf j^\top\mathbf D_0^{-1}\mathbf j}{p_0(\mathbf x_i)}
    \,\mathrm{d}\mathbf x_i,
    \label{eq:single_EPR}
\end{equation}
is the EPR of an individual subsystem. Its stationary probability distribution is Gaussian \citep{gardiner2004handbook},
\begin{equation}
    p_0(\mathbf x_i)
    =
    \frac{1}{(2\pi)^{d/2}\sqrt{\det\mathbf S_0}}
    \exp\left[
        -\frac{1}{2}\mathbf x_i^\top\mathbf S_0^{-1}\mathbf x_i
    \right],
    \label{eq:Gaussian_dist}
\end{equation}
where the covariance matrix $\mathbf S_0$ satisfies the Lyapunov equation
\begin{equation}
    \mathbf M_0\mathbf S_0+\mathbf S_0\mathbf M_0^\top
    =
    -2\mathbf D_0.
    \label{eq:uncoupled-lyapunov}
\end{equation}
The stationary probability current is $\mathbf j_0(\mathbf x_i)=\mathbf Q_0\mathbf S_0^{-1}\mathbf x_i\,p_0(\mathbf x_i)$, where the antisymmetric irreversibility matrix
\begin{equation}
    \mathbf Q_0
    =
    \mathbf M_0\mathbf S_0+\mathbf D_0,
    \label{eq:def_Q_matrix}
\end{equation}
is a measure of stationary irreversible circulation in the system. The EPR can therefore also be written as \citep{Godreche2019Characterising}
\begin{equation}
    \begin{split}
        \Phi_1
        &=
        \operatorname{tr}\left[
            \mathbf S_0^{-1}\mathbf Q_0^\top
            \mathbf D_0^{-1}\mathbf Q_0
        \right].
    \end{split}
    \label{eq:EPR_Q}
\end{equation}
This expression is non-negative and vanishes if and only if $\mathbf Q_0=\mathbf 0$, corresponding to detailed balance.

\subsection{Coupling matrix}
We couple the subsystems in Eq.~(\ref{eq:langevin}) through a generic coupling matrix of the form
\begin{equation}
    \mathbf \Theta(\omega_c)
    =
    -\omega_c\sum_{l\in\mathcal C}\mathbf K_l\otimes\mathbf G_l,
    \label{generic_coupling_matrix}
\end{equation}
where $\mathcal C\subseteq\{1,\dots,d\}$ is the set of coupled coordinates, $\mathbf K_l\in\mathbb R^{n\times n}$ is a symmetric positive-semidefinite topology matrix, $\mathbf G_l\in\mathbb R^{d\times d}$ selects coordinate $l$, and $\omega_c>0$ is the coupling strength. For example, diffusive coupling between the $y$ coordinates of two two-dimensional subsystems, $\mathbf x_i=(x_i,y_i)^\top$, is described by
\begin{equation}
    \mathbf K_y=
    \begin{pmatrix}
        1 & -1\\
        -1 & 1
    \end{pmatrix},
    \qquad
    \mathbf G_y=
    \begin{pmatrix}
        0 & 0\\
        0 & 1
    \end{pmatrix},
\end{equation}
so that $\mathbf\Theta=-\omega_c\mathbf K_y\otimes\mathbf G_y$ contributes forces $-\omega_c(y_1-y_2)$ and $-\omega_c(y_2-y_1)$ to the respective $y$ equations, while leaving the $x$ equations unchanged.

For the general coupling matrix in Eq.~(\ref{generic_coupling_matrix}), the coupled dynamics read
\begin{equation}
    \frac{\mathrm d\mathbf x(t)}{\mathrm dt}
    =
    \mathbf M\mathbf x(t)
    +
    \sqrt{2\mathbf D}\,\boldsymbol{\eta}(t),
    \label{eq:coupled-dynamics}
\end{equation}
where $\mathbf M=\boldsymbol{\Gamma}^{-1}(\boldsymbol{\Lambda}+\mathbf\Theta)$ and $\mathbf D=\mathbf I_n\otimes\mathbf D_0=\boldsymbol{\Gamma}^{-1}\mathbf T^{\mathrm{eff}}$, with $\mathbf x=(\mathbf x_1^\top,\dots,\mathbf x_n^\top)^\top$ and $\boldsymbol{\eta}=(\boldsymbol{\eta}_1^\top,\dots,\boldsymbol{\eta}_n^\top)^\top$. Here, $\boldsymbol{\Gamma}=\mathbf I_n\otimes\boldsymbol{\Gamma}_0$, $\boldsymbol{\Lambda}=\mathbf I_n\otimes\boldsymbol{\Lambda}_0$, and $\mathbf T^{\mathrm{eff}}=\mathbf I_n\otimes\mathbf T_0^{\mathrm{eff}}$. 

\subsection{Modal decomposition}
\label{subsection:modal_decomp}
For the following calculations, we assume that the topology
matrices in Eq.~(\ref{generic_coupling_matrix}) commute
pairwise, $[\mathbf K_l,\mathbf K_{l'}]=0$ for all
$l,l'\in\mathcal C$. Since these matrices are symmetric,
they can be simultaneously diagonalised. In their common
eigenbasis, the system decomposes into independent modes,
and the drift matrix takes the block-diagonal form
\begin{equation}
    \mathbf M
    =
    \bigoplus_{m=1}^n\mathbf M_m
    =
    \bigoplus_{m=1}^n
    \boldsymbol{\Gamma}_0^{-1}
    (\boldsymbol\Lambda_0+\omega_c\boldsymbol\kappa_m),
    \label{eq:m-decomposed}
\end{equation}
where $\boldsymbol\kappa_m=-\sum_{l\in\mathcal C}\lambda_{l,m}\mathbf G_l$ \citep{Pecora1998MasterStability}. Here, $\lambda_{l,m}$ is the eigenvalue of $\mathbf K_l$ in mode $m$.

For simplicity, we assume that the topology matrices have the same
zero-eigenvalue modes, such that for each mode $m$, either
$\lambda_{l,m}=0$ for all $l\in\mathcal C$, or
$\lambda_{l,m}>0$ for all $l\in\mathcal C$. This assumption is not essential and is made only to avoid defining the coupled and residual variables separately for each mode in Sec.~\ref{subsection:coupled-residual-variables}.

To ensure that a steady state exists, we assume that
$\mathbf M_m(\omega_c)$ is Hurwitz stable for all $\omega_c\geq 0$. Therefore, the total EPR can be written as the sum of the EPR contributions from the individual modes,
\begin{equation}
    \Phi_n(\omega_c)=\sum_{m=1}^n\tilde{\Phi}_m(\omega_c),
    \label{eq:modal-epr-sum}
\end{equation}
where $\tilde{\Phi}_m(\omega_c)$ denotes the EPR contribution of mode $m$, and for each mode, the covariance matrix $\mathbf S_m$ solves the $d$-dimensional Lyapunov equation with drift $\mathbf M_m(\omega_c)$. Details of this modal decomposition are given in Appendix~\ref{app:modal-decomposition}. 

For network topologies with collective zero modes, the coupling does not contribute to these modes, since $\boldsymbol\kappa_m=\mathbf{0}$. Consequently, the dynamics of the collective modes are unaffected by coupling, and their steady-state covariance and EPR remain identical to those of an uncoupled subsystem.

\begin{figure}
    \centering
    \includegraphics[width=\linewidth]{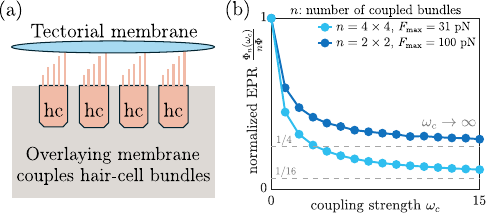}
    \caption{Coupling reduces EPR in a model of bullfrog hair-cell bundles.
(a) Schematic of an array of hair-cell (hc) bundles mechanically coupled through an overlying elastic membrane. The membrane couples the bundle-position coordinates.
(b) Normalized steady-state EPR as a function of the membrane coupling strength $\omega_c$ in the quiescent fixed-point regime at $F_{\max}=31\,\mathrm{pN}$ and $F_{\max}=100\,\mathrm{pN}$. All remaining single-bundle parameters are those of the bullfrog hair-bundle model in Ref.~\citep{hair_cell_Roldan_2021}. The EPR decreases monotonically with increasing coupling and approaches a strong-coupling value close to $1/n$, consistent with the prediction of the linear theory in Eq.~(\ref{eq:modal_epr_2sys}).}
    \label{fig:panel2}
\end{figure}

\section{Two-dimensional systems ($d=2$)}
\label{section:2d}
We consider identical two-dimensional subsystems and show that increasing the coupling strength suppresses irreversible circulation in every nonzero coupling mode, thereby monotonically decreasing the corresponding modal EPR, as illustrated in Fig.~\ref{fig:panel1}(a) and (b).
\subsection{Coupling reduces EPR monotonically}

For a two-dimensional subsystem, we take $\mathbf{x}_i=(x_i,y_i)^\top$, $\boldsymbol{\Gamma}_0=\mathbf I_2$, and
\begin{equation}
\boldsymbol \Lambda _0 =
\begin{pmatrix}
-a & -b \\
c & -d
\end{pmatrix},
\qquad
\mathbf{D}_0 =
\begin{pmatrix}
D_x & 0 \\
0 & D_y
\end{pmatrix},
\label{eq:2D-setup}
\end{equation}
where $a,b,c,d\in\mathbb R$ and $D_x,D_y>0$. Stability of the uncoupled system requires $a+d>0$ and $ad+bc>0$, while broken detailed balance requires $cD_x+bD_y\neq0$. We couple the $y$ coordinates through the coupling matrix $\mathbf \Theta(\omega_c)=-\omega_c\mathbf K_y\otimes\mathbf G_y$, where $\mathbf G_y=\mathrm{diag}(0,1)$ and the positive-semidefinite matrix $\mathbf K_y$ specifies the coupling network topology. Here, we restrict $a$ to be positive to ensure that the coupled system remains stable for all $\omega_c\geq0$.

The EPR contribution of mode $m$ depends only on the corresponding eigenvalue $\lambda_{y,m} \equiv\lambda_m$ of $\mathbf K_y$ and is
\begin{equation}
\tilde{\Phi}_m(\omega_c)=
\frac{(cD_x+bD_y)^2}
{D_xD_y\left(a+d+\lambda_m\omega_c\right)}.
\label{eq:modal_epr_2sys}
\end{equation}

Hence, $\tilde{\Phi}_m$ decreases monotonically with $\omega_c$
for every mode with $\lambda_m>0$, while any collective modes
with $\lambda_m=0$ remain unchanged. This decrease is due to the monotonic suppression of irreversible circulation, as
characterized by the irreversibility matrix $\mathbf Q_m$
(see Appendix~\ref{app:d2-current-suppression}).
In particular, for every mode with $\lambda_m>0$, as
$\omega_c\to\infty$, fluctuations along the coupled coordinate
vanish and $\mathbf S_m$ approaches a rank-deficient limit, with
\begin{subequations}
\begin{align}
    [\mathbf S_m]_{x,y} &\to 0, \\
    [\mathbf S_m]_{y,y} &\to 0 .
\end{align}
\end{subequations}
As illustrated in Fig.~\ref{fig:panel1}(b), increasing coupling confines the probability current to a narrower region of phase space. In the strong-coupling limit, we also obtain $\|\mathbf j_m\|_{L^1}\to0$ for every mode with $\lambda_m>0$, as shown in Appendix~\ref{app:d2-current-suppression}.
To examine whether the resulting reduction in EPR persists beyond the linear setting, we next consider an array of coupled vertebrate hair-cell bundles.

\subsection{Biological example: 2D model of hair-cell bundles} 

Vertebrate hair cells in the inner ear convert mechanical deflections of hair bundles into electrical signals and amplify weak inputs over a broad dynamic range \citep{hair_cell_Roldan_2021,Gillespie2009,hair_cell_julicher,Martin2001,Hudspeth1997}. Neighboring bundles are mechanically coupled by an overlying tectorial membrane (see Fig.~\ref{fig:panel2}(a)), which can enhance their collective response \citep{Robles2001,Freeman2003,Dierkes2008Enhancement}. To test our result in a nonlinear, spatially extended system, we consider an array of $n$ identical hair bundles on a square lattice. The governing Langevin equations read
\begin{subequations}
\begin{align}
\lambda_x\dot{X}_{(i,j)}
&=
-\frac{\partial V}{\partial X_{(i,j)}}
-\frac{\partial U_{\mathrm{mem}}}{\partial X_{(i,j)}}
+\sqrt{2k_{\mathrm B}T\lambda_x}\,\eta_x^{(i,j)},\\
\lambda_c\dot{C}_{(i,j)}
&=
-\frac{\partial V}{\partial C_{(i,j)}}
-F_{\mathrm{act}}
+\sqrt{2k_{\mathrm B}T_{\mathrm{eff}}\lambda_c}\,\eta_c^{(i,j)}.
\end{align}
\label{eq:hair-cell}
\end{subequations}
Here, each bundle is described by two variables: its position $X_{(i,j)}$ and an effective molecular-motor coordinate $C_{(i,j)}$ \citep{hair_cell_Roldan_2021,Dierkes2008Enhancement}. The potential $V(X_{(i,j)},C_{(i,j)})$ describes the elastic elements of the hair bundle and the mechano-sensitive ion channels, $U_{\mathrm{mem}}$ is the membrane-interaction potential, $F_{\mathrm{act}}$ is the active motor force, $\lambda_x$ and $\lambda_c$ are friction coefficients, and $\eta_{x}^{(i,j)}, \eta_{c}^{(i,j)}$ represent unit white noise terms. We assume that the overlying membrane couples the bundle positions through elastic interactions as in Ref.~\cite{Dierkes2012} and that the deflections occur only along one direction. Further details of the model and membrane-mediated coupling are given in Appendix~\ref{app:hair-cell-model}.

Since the force derived from the potential $V$ in Eq.~(\ref{eq:hair-cell}) is nonlinear, the EPR is estimated numerically from simulated trajectories using the framework in Ref.~\citep{Sekimoto1998LangevinThermodynamics}, as described in Appendix~\ref{AppendixTrajEPR}. Using the parameters of the bullfrog hair-bundle model in Ref.~\citep{hair_cell_Roldan_2021}, we find that at small active forces, $F_{\max}\lesssim 40\,\mathrm{pN}$, fluctuations remain close to the stable fixed point and nonlinear corrections are weak. In this regime, the normalized EPR decreases with increasing coupling and approaches a strong-coupling value close to the linear-theory prediction of $1/n$, as shown in Fig.~\ref{fig:panel2}(b). At intermediate forces, nonlinear effects become important and the strong-coupling EPR can differ substantially from the linear prediction (see Appendix~\ref{app:hair-cell-model}). At large forces, $F_{\max}\gtrsim 95\,\mathrm{pN}$, fluctuations again remain close to a stable fixed point, where the dynamics are locally linear, and the normalized strong-coupling EPR approaches values close to $1/n$. Since the absolute EPR is much larger in this regime than in the weak-force regime \citep{hair_cell_Roldan_2021}, the absolute reduction in EPR due to coupling is particularly large.

\section{Higher-dimensional systems ($d>2$)}

The monotonic decrease of EPR with coupling is guaranteed only in two-dimensional systems. In higher-dimensional systems, the EPR may increase or decrease with coupling strength and can vary non-monotonically. Nevertheless, as the coupling strength increases, the irreversible-current components associated with the coupled DoF are asymptotically suppressed, as shown in Fig.~\ref{fig:panel3}(a). The residual DoF, however, can retain irreversible circulation. Consequently, the EPR in the strong-coupling limit need not be smaller than the EPR of the uncoupled system. Whether coupling eventually leads to lower dissipation than in the uncoupled system is determined by the EPR of the residual dynamics that survives in the strong-coupling limit.

\subsection{Coupled and residual variables}
\label{subsection:coupled-residual-variables}
To characterize the strong-coupling behavior in higher dimensions, we separate the DoF of each nonzero coupling mode $m$ into directly coupled variables $\mathbf u_m\in\mathbb R^k$ and residual variables $\mathbf v_m\in\mathbb R^{d-k}$. When writing variables in an ordered vector $(\mathbf u_m,\mathbf v_m)$, the diffusion matrix is $\mathbf D_0=\operatorname{diag}(\mathbf D_u,\mathbf D_v)$, and the modal drift matrix in Eq.~(\ref{eq:m-decomposed}) assumes a block form
\begin{equation}
    \mathbf M_m(\omega_c)=
    \begin{pmatrix}
        \omega_c\mathbf A_0+\mathbf A & \mathbf B\\
        \mathbf C & \mathbf R
    \end{pmatrix},
    \label{eq:modal-block-drift-main}
\end{equation}
where $\mathbf A$, $\mathbf B$, $\mathbf C$, and $\mathbf R$ are the blocks corresponding to the drift $\boldsymbol\Gamma_0^{-1}\boldsymbol\Lambda_0$, and $\mathbf A_0$ is the coupled-variable block of $\boldsymbol\Gamma_0^{-1}\boldsymbol\kappa_m$. The mode index on these block matrices is suppressed for notational simplicity. We require $\mathbf R$ to be Hurwitz stable so that Eq.~(\ref{eq:residual-lyapunov-main}), derived below, admits a unique stationary covariance solution.

\subsection{Suppression of currents along the coupled DoF}

For each nonzero coupling mode $m$, the covariance matrix can be written in block form as
\begin{equation}
    \mathbf{S}_m =
    \begin{pmatrix}
        \mathbf{S}_{uu} & \mathbf{S}_{uv} \\
        \mathbf{S}_{uv}^{\top} & \mathbf{S}_{vv}
    \end{pmatrix},
    \label{eq:covariance-block-main}
\end{equation}
whose blocks have the asymptotic form
\begin{subequations}
    \begin{align}
        \mathbf{S}_{uu} &= \mathcal{O}(\omega_c^{-1}),\\
        \mathbf{S}_{uv} &= \mathcal{O}(\omega_c^{-1}),\\
        \mathbf{S}_{vv} &= \mathbf{Z}_0+\mathcal{O}(\omega_c^{-1}),
    \end{align}
    \label{eq:covariance-scaling-main}
\end{subequations}
in the strong-coupling limit, where $\mathbf{Z}_0$ is the covariance matrix which obeys the Lyapunov equation
\begin{equation}
    \mathbf{R}\mathbf{Z}_0
    +
    \mathbf{Z}_0\mathbf{R}^{\top}
    =
    -2\mathbf{D}_v,
    \label{eq:residual-lyapunov-main}
\end{equation}
with drift $\mathbf{R}$ and diffusion $\mathbf{D}_v$. Thus, fluctuations along the coupled DoF and their correlations with the residual DoF vanish in the strong-coupling limit, while the fluctuations of the residual system are described by Eq.~(\ref{eq:residual-lyapunov-main}). A proof of this convergence is given in Appendix~\ref{app:covariance-expansion}. 

Related results have been obtained for stochastic systems with separated time scales, particularly in the context of eliminating fast degrees of freedom~\cite{BoCelani2014,Vishen2020,ChunNoh2015,Busiello2020,Wang2016}. In the present setting, the coupling strength modifies only the drift matrix of the coupled variables, while the diffusion matrix and their interactions with the residual DoF remain independent of $\omega_c$.

Analogous to Eq.~(\ref{eq:covariance-block-main}), the irreversibility matrix $\mathbf{Q}_m$ can also be partitioned into coupled and residual blocks. Using the covariance expansion in Eq.~(\ref{eq:covariance-scaling-main}), the corresponding blocks have the asymptotic form (see Appendix~\ref{app:covariance-expansion})
\begin{subequations}
    \begin{align}
        \mathbf{Q}_{uu} &= \mathcal{O}(\omega_c^{-1}),\\
        \mathbf{Q}_{uv} &= \mathcal{O}(\omega_c^{-1}),\\
        \mathbf{Q}_{vv} &= \mathbf{R}\mathbf{Z}_0 + \mathbf{D}_v + \mathcal{O}(\omega_c^{-1}).
    \end{align}
    \label{eq:Q-scaling-main}
\end{subequations}
Consequently, the irreversible circulation in coupled DoF vanishes in the strong coupling limit.
The component of the stationary probability current of mode $m$ along the coupled DoF, $\mathbf j_u$, also vanishes in this limit, since (see Appendix~\ref{app:residual-convergence})
\begin{align}
\|\mathbf{j}_u\|_{L^1}^2
&\leq
\operatorname{tr}
\left[
\mathbf{Q}_u\mathbf{S}_m^{-1}\mathbf{Q}_u^\top
\right] \to 0,
\label{eq:ju-L1-bound}
\end{align}
where $\mathbf{Q}_u=(\mathbf{Q}_{uu},\mathbf{Q}_{uv})$. In systems with $d\geq 3$, however, the residual component $\mathbf j_v$ can remain nonzero, and therefore the full modal probability current $\mathbf j_m$ need not vanish.

\subsection{Residual dynamics determine the strong-coupling limit}

The EPR contribution of each nonzero coupling mode $m$ can be decomposed as
\begin{equation}
    \tilde{\Phi}_m(\omega_c)
    =
    \tilde{\Phi}_u(\omega_c)
    +
    \tilde{\Phi}_v(\omega_c),
\end{equation}
where $\tilde{\Phi}_u,\tilde{\Phi}_v \geq 0$ are the EPR contributions associated with the coupled and residual DoF, respectively.

The strong-coupling asymptotics in Eqs.~(\ref{eq:covariance-scaling-main}) and (\ref{eq:Q-scaling-main}) imply that the contribution associated with the coupled DoF vanishes as $\omega_c \to \infty$,
\begin{equation}
    \tilde{\Phi}_u(\omega_c)
    \to 0,
    \label{eq:phi-u-limit-main}
\end{equation}
while the contribution associated with the residual DoF converges to
\begin{equation}
    \tilde{\Phi}_v(\omega_c)
    \to
    \tilde{\Phi}_{\rm res},
    \label{eq:phi-v-limit-main}
\end{equation}
where $\tilde{\Phi}_{\rm res}\equiv\Phi(\mathbf{R},\mathbf{D}_v)$ is the EPR of the residual dynamics with drift $\mathbf{R}$ and diffusion $\mathbf{D}_v$, whose stationary covariance $\mathbf Z_0$ is determined by Eq.~(\ref{eq:residual-lyapunov-main}). Consequently, in the limit $\omega_c\to\infty$, the EPR contribution of mode $m$ converges to
\begin{equation}
    \tilde{\Phi}_m(\omega_c)
    \to
    \tilde{\Phi}_{\rm res},
    \label{eq:modal-residual-limit-main}
\end{equation}
as shown in Appendix~\ref{app:residual-convergence}.

The residual EPR $\tilde{\Phi}_{\rm res}$ therefore determines whether the EPR of a nonzero mode in the strong-coupling limit is smaller than its uncoupled value. For a given mode, this occurs when $\tilde{\Phi}_{\rm res}<\tilde{\Phi}_m(0)$.

For the full system, let $n_0$ denote the number of zero-eigenvalue modes and label them by $m=1,\ldots,n_0$. These modes are unaffected by the coupling, whereas the EPR of each nonzero mode converges to its corresponding residual EPR $\tilde{\Phi}_{{\rm res},m}$. It follows that the strongly coupled system dissipates less than the uncoupled system if
\begin{equation}
    n_0\Phi_1
    +
    \sum_{m=n_0+1}^{n}
    \tilde{\Phi}_{{\rm res},m}
    <
    n\Phi_1,
    \label{eq:residual_advantage}
\end{equation}
where $\Phi_1$ is the EPR of one uncoupled subsystem. Equation~(\ref{eq:residual_advantage}) therefore provides the criterion for whether coupling ultimately leads to lower dissipation than in the uncoupled system. The convergence to the residual dynamics, together with this criterion, constitutes the central result of this work. This criterion is sufficient but not necessary for coupling to reduce dissipation at finite coupling strengths. Since the EPR can vary non-monotonically with coupling strength, an energetic advantage may also arise at weak or intermediate coupling, even when the strong-coupling limit exceeds the uncoupled EPR.

The convergence of the covariance in Eq.~(\ref{eq:covariance-scaling-main}) does not by itself guarantee convergence of the EPR. Since the covariance becomes singular along the coupled DoF, the corresponding components of its inverse grow with $\omega_c$, as shown in Eq.~(\ref{eq:inverse-covariance-scaling}). The convergence in Eq.~(\ref{eq:modal-residual-limit-main}) therefore also relies on the suppression of the irreversible-current components in Eq.~(\ref{eq:Q-scaling-main}), which ensures that their contribution to the EPR vanishes. As we show in Sec.~\ref{subsection:main-different-temp}, this suppression can fail when the directly coupled DoF have different effective temperatures, even though the covariance still converges.

For the identical two-dimensional subsystems considered in Sec.~\ref{section:2d}, the residual dynamics is one-dimensional and therefore $\tilde{\Phi}_{\rm res}=0$, while Eq.~(\ref{eq:modal_epr_2sys}) independently guarantees that the EPR decreases monotonically with coupling. In higher dimensions, by contrast, the EPR need not vary monotonically with coupling. Nevertheless, Eq.~(\ref{eq:modal-residual-limit-main}) shows that the EPR of each nonzero mode converges in the strong-coupling limit to its residual value $\tilde{\Phi}_{\rm res}$, which need not vanish.

\subsection{EPR trend near strong-coupling limit}

To characterize how the modal EPR approaches its strong-coupling limit, $\omega_c \to \infty$, we consider the asymptotic expansions of the contributions associated with the coupled and residual DoF,
\begin{subequations}
\begin{align}
\tilde{\Phi}_u(\omega_c)
&=
\frac{K_u}{\omega_c}
+\mathcal{O}(\omega_c^{-2}),
\label{phi_u}
\\
\tilde{\Phi}_v(\omega_c)
&=
\tilde{\Phi}_{\rm res}
+\frac{K_v+\tilde K}{\omega_c}
+\mathcal{O}(\omega_c^{-2}),
\label{phi_v}
\end{align}
\end{subequations}
where $K_u$, $K_v$, and $\tilde K$ are the first-order expansion coefficients associated with these EPR contributions. Consequently, the modal EPR approaches its residual value as
\begin{equation}
    \tilde{\Phi}_m(\omega_c)
    =
    \tilde{\Phi}_{\rm res}
    +
    \frac{K}{\omega_c}
    +
    \mathcal{O}(\omega_c^{-2}),
    \label{eq:modalEPR_expansion}
\end{equation}
where $K=K_u+K_v+\tilde K$.

The coefficient $K_u$ gives the leading correction to $\tilde{\Phi}_u$ and originates from $\mathbf Q_{uu}$, which represents the remaining irreversible circulation within the coupled variables at large but finite coupling. For the residual variables, $K_v$ arises from current contributions involving $\mathbf Q_{vu}$, whereas $\tilde K$ captures the leading correction to the EPR associated with the residual circulation. We show in Appendix~\ref{app:first-order-coefficient} that the coefficients $K_u$ and $K_v$ are non-negative. Physically, both these coefficients are associated with probability-current contributions involving the coupled variables, which are suppressed as the strong-coupling limit is approached, consistent with their non-negative contribution to the finite-coupling EPR correction. Consequently, $K_u$ and $K_v$ can only contribute toward an asymptotic decrease of the EPR with increasing coupling. In contrast, $\tilde K$ can have either sign and therefore provides the only contribution that can lead to an increase of the EPR as $\omega_c \to \infty$. Since $K_u,K_v\geq0$, for $K\neq0$ the modal EPR increases asymptotically if and only if $\tilde K<-(K_u+K_v)$. At $\tilde K=-(K_u+K_v)$, the first-order correction vanishes and higher-order terms determine the trend.

To understand the physical origin of the sign-indefinite correction $\tilde K/\omega_c$, we introduce the normalized irreversible-current matrix for the residual variables, $\mathbf J_{vv}=\mathbf D_v^{-1/2}\mathbf Q_{vv}\mathbf S_{vv}^{-1/2}$, which can be expanded around the residual dynamics as
\begin{equation}
    \mathbf J_{vv}
    =
    \mathbf J_{\rm res}
    +
    \frac{\mathbf J_1}{\omega_c}
    +
    \mathcal O(\omega_c^{-2}).
    \label{eq:J_expansion}
\end{equation}
Here $\mathbf J_{\rm res}$ is the normalized irreversible-current matrix of the residual dynamics, satisfying $\tilde{\Phi}_{\rm res}=\|\mathbf J_{\rm res}\|_F^2$, while $\mathbf J_1$ is the leading finite-coupling correction due to the coupled DoF. The coefficient $\tilde K$ can then be expressed as
\begin{equation}
    \tilde K
    =
    2\langle \mathbf J_{\rm res},\mathbf J_1\rangle_F.
    \label{eq:inner-product}
\end{equation}
The complete derivation is given in Appendix~\ref{app:current-alignment}.

Physically, the coefficient $\tilde K$ measures whether the correction $\mathbf J_1$ reinforces or screens the residual circulation $\mathbf J_{\rm res}$. This representation provides a mechanistic interpretation of how, near the strong-coupling limit, the circulation of the residual DoF changes as the coupling strength is increased, and how this change contributes to the corresponding EPR trend. For $\tilde K>0$, the correction is aligned with the residual current, so the remaining influence of the coupled DoF reinforces the irreversible circulation in the residual variables. Increasing $\omega_c$ suppresses this reinforcing contribution, thereby lowering the corresponding EPR contribution. For $\tilde K<0$, the correction instead opposes the residual current, so the remaining influence of the coupled DoF partially screens the residual circulation. Increasing $\omega_c$ removes this screening contribution and can therefore increase the EPR as the residual limit is approached.

If $\tilde{\Phi}_{\rm res}<\tilde{\Phi}_m(0)$ and $K>0$, there exists a coupling threshold beyond which increasing the coupling strength decreases the modal EPR monotonically toward a value below its uncoupled value. If instead $\tilde{\Phi}_{\rm res}<\tilde{\Phi}_m(0)$ and $K<0$, the modal EPR approaches the residual value from below, implying the existence of at least one optimal finite coupling strength at which the EPR is minimized.

Furthermore, coupling $d-1$ variables leaves a one-dimensional residual system, which, if stable on $\mathbb R$, cannot sustain a stationary irreversible current. Hence, $\mathbf J_{\rm res}=0$, so that $\tilde{\Phi}_{\rm res}=0$ and $\tilde K=0$. The same statement holds whenever the residual dynamics satisfies detailed balance. In both cases, the sign-indefinite contribution vanishes, and the asymptotic coefficient $K$ is determined entirely by the non-negative contributions $K_u$ and $K_v$. The $d=2$ case is the simplest example of this behavior.

\begin{figure}
    \centering
    \includegraphics[width=\linewidth]{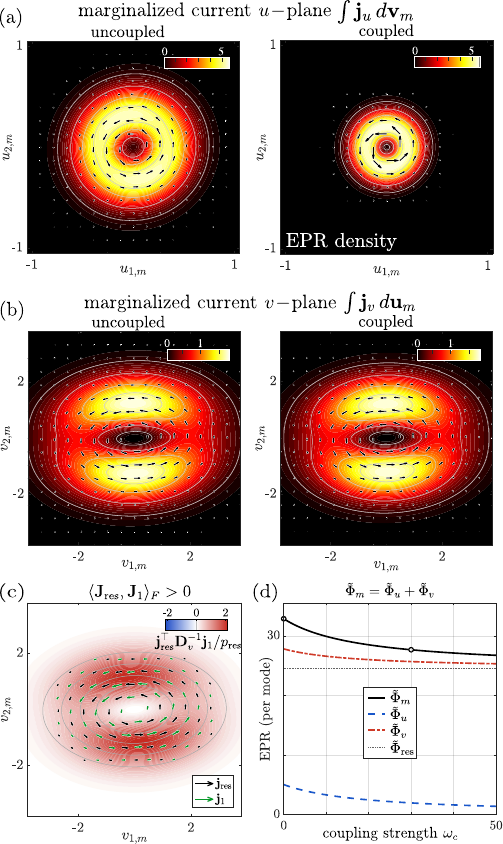}
    \caption{
    Mechanism of asymptotic EPR reduction in a representative higher-dimensional mode.
    We consider $d=4$ and $k=2$, with directly coupled DoF
    $\mathbf u_m=(u_{1,m},u_{2,m})^\top$ and residual DoF
    $\mathbf v_m=(v_{1,m},v_{2,m})^\top$.
    (a) Marginalized probability current in the coupled DoF,
    $\int \mathbf j_u\,d\mathbf v_m$ (arrows), shown over the projected EPR
    density
    $\bar{\sigma}_u=\int \mathbf j_u^\top\mathbf D_u^{-1}\mathbf j_u/p\,
    d\mathbf v_m$, at $\omega_c=0$ and $\omega_c=30$.
    Increasing coupling strongly suppresses both the current and its EPR
    contribution $\tilde{\Phi}_u$.
    (b) Marginalized current in the residual DoF,
    $\int \mathbf j_v\,d\mathbf u_m$, shown over
    $\bar{\sigma}_v=\int \mathbf j_v^\top\mathbf D_v^{-1}\mathbf j_v/p\,
    d\mathbf u_m$. A finite irreversible circulation remains at strong
    coupling and converges to the residual current.
    (c) Alignment between the fields $\mathbf j_{\rm res}$ and
$\mathbf j_1$, which are associated with the residual
current $\mathbf J_{\rm res}$ and its leading finite-coupling
correction $\mathbf J_1$, respectively. The background shows the signed overlap density
    $\mathbf j_{\rm res}^\top\mathbf D_v^{-1}\mathbf j_1/p_{\rm res}$,
    whose integral is $\langle\mathbf J_{\rm res},\mathbf J_1\rangle_F$.
    For the parameters shown, the positive overlap indicates that the
    finite-coupling correction reinforces the residual circulation.
    (d) Modal EPR $\tilde{\Phi}_m$ as a function of $\omega_c$, approaching
    $\tilde{\Phi}_{\rm res}$ from above. The monotonic decrease in this
    example allows the currents and their finite-coupling correction in
    (a)--(c) to directly illustrate the asymptotic mechanism.
    \textit{Simulation parameters:}
    $a=20$, $p=7$, $r_1=0.5$, $r_2=1.5$, $q_1=5$, $q_2=2$, and
    $b_1=b_2=c_1=c_2=1.5$.
    }
    \label{fig:panel3}
\end{figure}
\subsection{An exemplary model ($d=4$)}
To elaborate on the above mechanistic discussion, we consider a single mode of a representative four-dimensional coupled system, comprising two coupled variables $\mathbf u=(u_1,u_2)^\top$ and two residual variables $\mathbf v=(v_1,v_2)^\top$. In this example, we take $\mathbf A_0=-\mathbf I_2$, $\mathbf B=\mathrm{diag}(b_1,b_2)$, $\mathbf C=\mathrm{diag}(c_1,c_2)$, and $\mathbf D_u=\mathbf D_v=\mathbf I_2$, with
\begin{equation}
\mathbf A=
\begin{pmatrix}
-a & -p\\
p & -a
\end{pmatrix},
\qquad
\mathbf R=
\begin{pmatrix}
-r_1 & -q_1\\
q_2 & -r_2
\end{pmatrix}.
\label{eq:representative_example}
\end{equation}
We take $a,r_1,r_2,q_1,q_2,p>0$, so that the diagonal blocks
$\mathbf A$ and $\mathbf R$ are Hurwitz, and restrict the remaining
parameters such that $\mathbf M_m(\omega_c)$ remains Hurwitz for
$\omega_c\geq0$. In this example, the residual dynamics are two-dimensional, with EPR as
\begin{equation}
    \tilde\Phi_{\text{res}}=\frac{q^2}{r},
\end{equation}
where $q\equiv q_1+q_2$ and $r\equiv r_1+r_2$. Further, for large but finite coupling, the leading quasi-static
deterministic response of the coupled variables is $\mathbf u\simeq-\frac{1}{\omega_c}\mathbf A_0^{-1}\mathbf B\mathbf v$, and consequently the dynamics of the residual variables are approximately
\begin{equation}
    \dot{\mathbf v}\simeq
    \left(
    \mathbf R+\frac{\mathbf C\mathbf B}{\omega_c}
    \right) \mathbf v+ \sqrt{2}\,\boldsymbol{\eta}_v,
\end{equation}
where $\boldsymbol{\eta}_v$ is the noise in the residual dynamics. To see how this affects the irreversible circulation, we consider the irreversibility matrix $\mathbf Q_{vv}$, which for large $\omega_c$ reads
\begin{equation}
    \mathbf Q_{vv}\simeq\mathbf Q_{\text{res}}+\frac{1}{\omega_c}\frac{q(b_1c_1+b_2c_2)}{r^2}\boldsymbol{\Omega}
    =\mathbf Q_{\text{res}}+\frac{1}{\omega_c}\mathbf Q_1,
\end{equation}
where $\boldsymbol{\Omega}=\begin{pmatrix}0&-1\\1&0\end{pmatrix}$ and $\mathbf Q_{\text{res}}=\frac{q}{r}\boldsymbol{\Omega}$. Here, for a fixed $\omega_c$, if $b_ic_i>0$ for both $i=1,2$, then the damping in the residual system is effectively reduced. As a result, the rotational components become relatively larger, and hence the net circulation in the system is increased. Therefore, as $\omega_c$ is increased, the circulation of the coupled variables is suppressed, as depicted in Fig.~\ref{fig:panel3}(a), and their contribution to $\mathbf Q_{vv}$ is correspondingly reduced. Consequently, the current amplitude in the residual variables decreases, as shown in Fig.~\ref{fig:panel3}(b), thereby contributing to a reduction in the EPR.

In addition to changing $\mathbf Q_{vv}$, the coupling also changes the
stationary covariance $\mathbf S_{vv}$ and hence the distribution over which
the irreversible current flows. The normalized current matrices
$\mathbf J_{\text{res}}$ and $\mathbf J_1$ incorporate both effects and allow
the corresponding first-order EPR contribution to be written compactly.
For this example,
\begin{equation}
    \left\langle \mathbf J_{\text{res}},\mathbf J_1\right\rangle_F
    =
    \frac{q^2}{2r^2}(b_1c_1+b_2c_2).
\end{equation}
Thus, when $b_ic_i>0$ for both $i=1,2$, the first-order normalized-current
correction is positively aligned with the residual current, consistent with
the reinforcing circulation described above. Conversely, when the correction
opposes the residual current, the negative contribution can be sufficiently
large to reverse the asymptotic EPR trend. An explicit example for the same
model, with $\tilde K<-(K_u+K_v)$ and hence $K<0$, is given in
Appendix~\ref{app:epr-increase}.

In the example considered in Fig.~\ref{fig:panel3}, the parameters are chosen such that $\tilde\Phi_{\text{res}}<\tilde{\Phi}_m(0)$. For this case, the EPR varies monotonically with $\omega_c$ (see Fig.~\ref{fig:panel3}(d)). Consequently, the difference between the uncoupled and coupled cases
directly illustrates the asymptotic mechanism derived above, even though
the analytical expansion itself is valid only for large $\omega_c$. To visualize this normalized-current correction in the residual variables, we associate $\mathbf J_{\text{res}}$ and $\mathbf J_1$ with the corresponding vector fields with respect to the residual dynamics $\mathbf j_{\text{res}}(\mathbf v)= p_{\text{res}} \mathbf{D}_v^{1/2} \mathbf J_{\text{res}}\mathbf Z_0^{-1/2}\mathbf v $ and $ \mathbf j_1(\mathbf v)= p_{\text{res}}\,\mathbf{D}_v^{1/2}\mathbf J_1\mathbf Z_0^{-1/2}\mathbf v,$
where $\mathbf Z_0$ and $p_{\text{res}}$ are the covariance and stationary distribution of the residual dynamics. In Fig.~\ref{fig:panel3}(c), we display these currents and their overlap density, which, when integrated, yields the overlap
\begin{equation}
    \left\langle \mathbf J_{\text{res}},\mathbf J_1\right\rangle_F
    =\int\frac{\mathbf j_{\text{res}}^\top(\mathbf v) \mathbf{D}_v^{-1} \mathbf j_1(\mathbf v)}{p_{\text{res}}(\mathbf v)}
    \,d\mathbf v.
\end{equation}
Thus, regions in which $\mathbf j_1$ is aligned with $\mathbf j_{\text{res}}$ contribute positively to the first-order EPR correction, whereas regions in which the two currents oppose each other contribute negatively.

\section{Heterogeneous systems}

The results above were derived for identical subsystems, for which a decomposition into independent modes is possible. When the coupled subsystems have different internal parameters or effective temperatures, the corresponding modes are generally no longer independent. Differences in the internal dynamical parameters can alter the residual dynamics that survives at strong coupling and thereby alter both the limiting EPR and its asymptotic approach. Differences in the effective temperatures of the coupled variables have a distinctive effect, as the EPR need not converge to a finite residual value and can instead grow indefinitely with increasing coupling.

\subsection{Parameter heterogeneity changes residual dynamics}

We first consider heterogeneity in the internal dynamical parameters while keeping the friction and diffusion matrices identical between subsystems, such that we have a simplified setting in which differences in effective temperature are not introduced. As discussed in Sec.~\ref{subsection:modal_decomp}, the identical-subsystem structure allows the dynamics to be decomposed into independent coupling modes. With parameter heterogeneity, these modes are generally no longer independent, so the system can no longer be treated through the same modal decomposition. Nevertheless, the strong-coupling convergence underlying Eqs.~(\ref{phi_u}) and~(\ref{phi_v}) does not require a decomposition into independent modes or commuting coupling topologies. Instead, in the eigenbasis of the full coupling matrix in Eq.~(\ref{generic_coupling_matrix}), the coupled DoF associated with its nonzero eigenvalues can still be separated from the residual DoF associated with its kernel. This separation yields a block structure analogous to Eq.~(\ref{eq:modal-block-drift-main}), so that the convergence in the strong-coupling limit described by Eqs.~(\ref{phi_u}) and~(\ref{phi_v}) applies directly to the full heterogeneous system, even though the resulting variables no longer form independent modes. The coupled DoF can therefore still be suppressed, while the residual DoF converge to a finite residual dynamics.

As an illustration, we consider two diffusively coupled two-dimensional systems with
\begin{equation}
    \boldsymbol{\Lambda}_0^{(i)}
    =
    \begin{pmatrix}
        -a_i & -b_i\\
        c_i & -d_i
    \end{pmatrix},
    \qquad i=1,2,
\end{equation}
and coupling acting only on the second coordinate. We define
$\bar a=(a_1+a_2)/2$ and $\delta a=(a_1-a_2)/2$, with analogous definitions for $b,c,$ and $d$.
In collective and relative coordinates, reordered as $(y_r,x_c,y_c,x_r)$, we identify the directly coupled variable $\mathbf u=(y_r)$ and the residual variables $\mathbf v=(x_c,y_c,x_r)^\top$. Since the heterogeneous system does not decompose into independent modes, we omit the modal index $m$ used previously. The drift matrix then becomes
\begin{equation}
\begin{split}
    \mathbf M(\omega_c)
    &=
    \begin{pmatrix}
        -\bar d-2\omega_c & \delta c & -\delta d & \bar c\\
        -\delta b & -\bar a & -\bar b & -\delta a\\
        -\delta d & \bar c & -\bar d & \delta c\\
        -\bar b & -\delta a & -\delta b & -\bar a
    \end{pmatrix}\\
    &=
    \begin{pmatrix}
        \mathbf A+\omega_c\mathbf A_0 & \mathbf B\\
        \mathbf C & \mathbf R_{\rm het}
    \end{pmatrix}.
    \label{eq:heterogeneous-transformed}
\end{split}
\end{equation}
For identical subsystems, all $\delta$ terms vanish and the collective and relative modes decouple. Parameter heterogeneity mixes these modes, but only $y_r$ belongs to the coupled DoF, so its fluctuations and correlations are increasingly suppressed as $\omega_c$ grows and vanish as $\omega_c\to\infty$, while $(x_c,y_c,x_r)$ retain finite fluctuations. The surviving residual dynamics is described by
\begin{equation}
    \mathbf R_{\rm het}
    =
    \begin{pmatrix}
        -\bar a & -\bar b & -\delta a\\
        \bar c & -\bar d & \delta c\\
        -\delta a & -\delta b & -\bar a
    \end{pmatrix}.
    \label{eq:heterogeneous-residual}
\end{equation}

Therefore, differences in the internal dynamics can qualitatively alter the strong-coupling limit. In the case of identical systems, $x_r$ decouples from the collective dynamics and the residual part associated with the nonzero mode is one dimensional, so it cannot sustain an irreversible current of its own. When the internal parameters differ, $x_r$ remains coupled to the collective coordinates even after fluctuations in $y_r$ are suppressed. The residual system is then three-dimensional and can sustain additional irreversible circulation involving $x_r$, thereby modifying the residual EPR.

Accordingly, differences in the internal parameters modify the residual dynamics and can alter the limiting EPR in the strong-coupling regime. The strong-coupling expansion can again be written in the form of Eq.~(\ref{eq:modalEPR_expansion}), with $K$ denoting the leading coefficient for the full heterogeneous system. For sufficiently small parameter differences, the residual EPR, the uncoupled EPR, and $K$ remain close to their values in the corresponding identical system. Thus, if the identical system satisfies $\tilde{\Phi}_{\rm res}<\Phi_1$ and $K>0$, both inequalities persist under sufficiently small parameter perturbations, establishing local robustness of the asymptotic EPR reduction. The admissible range of heterogeneity is, however, model-dependent. Below, we illustrate this behavior in a model of coupled non-identical chemoreceptor signaling.

\subsection{Different effective temperatures}
\label{subsection:main-different-temp}

The effect of unequal effective temperatures on the strong-coupling limit depends on which variables are affected. When confined to the residual variables, this heterogeneity modifies the residual diffusion matrix $\mathbf D_v$, altering the residual steady state and its EPR. A qualitatively distinct behavior arises when directly coupled variables are maintained at different effective temperatures.

In the eigenbasis of the coupling matrix, differences in the effective temperatures of corresponding coupled variables generally result in off-diagonal diffusion terms involving the coupled coordinates. These terms introduce additional irreversible currents that depend on the temperature differences and persist in the strong-coupling limit. Consequently, the suppression of the irreversible current in the coupled DoF may not occur. As shown in Appendix~\ref{app:effective-temperatures}, the corresponding contribution to the EPR can scale as $\mathcal O(\omega_c)$ for $\omega_c\gg1$, so that the EPR grows with coupling strength and the convergence in Eq.~(\ref{eq:modalEPR_expansion}) no longer holds in general.

This behavior can be seen already for a pair of two-dimensional systems with identical drifts as in Eq.~(\ref{eq:2D-setup}), coupled diffusively through their $y$ coordinates, but with diffusion matrices
\begin{equation}
\mathbf D_0^{(i)}
=
\operatorname{diag}(D_x,D_y^{(i)}),
\qquad i=1,2.
\end{equation}
When $D_y^{(1)}\neq D_y^{(2)}$, the strong-coupling behavior is
\begin{equation}
    \Phi_2(\omega_c)
    =
    \frac{
        \left(D_y^{(1)}-D_y^{(2)}\right)^2
    }{
        2D_y^{(1)}D_y^{(2)}
    }\,
    \omega_c
    +
    \mathcal{O}(1).
    \label{eq:temperature-EPR}
\end{equation}
Physically, this linear contribution in $\omega_c$ arises from the coupling-mediated heat flux between variables maintained at different effective temperatures. This linear contribution in $\omega_c$ vanishes when the coupled variables have the same effective temperature, recovering convergence to the residual EPR for $\omega_c \to \infty$. 

More generally, when variables maintained at different effective temperatures are coupled, the stationary covariance still converges in the strong-coupling limit, with fluctuations along the directly coupled DoF vanishing. However, convergence of the covariance does not guarantee convergence of the EPR, since coupling mediates heat exchange between reservoirs at different effective temperatures, which can lead to increasing dissipation even as fluctuations vanish.

\subsection{Example: Coupling of heterogeneous bacterial chemoreceptors}
\begin{figure}
    \centering
    \includegraphics[width=\linewidth]{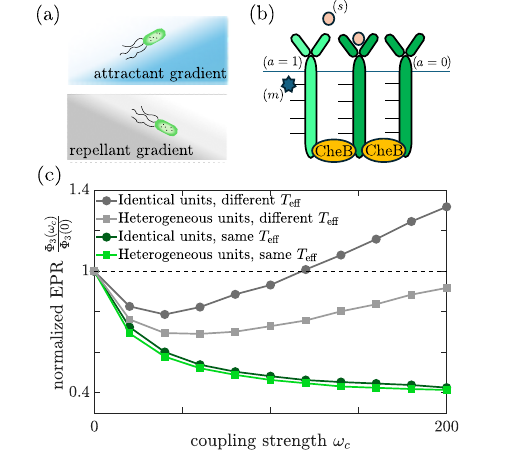}
   \caption{Dissipation in coupled heterogeneous chemotactic signaling units. (a) Schematic of chemotactic sensing along attractant and repellent gradients.
(b) Schematic of neighboring chemoreceptor units. Light and dark green indicate the active ($a=1$) and inactive ($a=0$) receptor states, respectively. Here $s$ denotes the ligand stimulus and $m$ the methylation level. Nearby units can interact through shared methylation and demethylation enzymes, CheR and CheB, which are coarse-grained in Eq.~(\ref{Eq:chemotaxis}) as diffusive coupling between their methylation variables.
(c) Normalized EPR as a function of coupling strength for three coupled nonlinear chemotactic signaling units with either identical or heterogeneous methylation rates. For equal effective temperatures of the methylation variables, the EPR decreases with coupling in both cases and approaches nearly the same strong-coupling value. When the coupled methylation variables have different effective temperatures, the EPR instead grows in the strong-coupling regime. \textit{Simulation parameters:} $\nu_a=50$, $s=10$, $m_0=4$, $\Delta_a=2$, and $a_0=0.5$.}
    \label{fig:panel5}
\end{figure}

Bacterial chemotaxis relies on an adaptive signaling network that allows cells to detect environmental chemical cues and bias their motion accordingly
\citep{adler1973chemotaxis_sugars, berg1977chemoreception}, as depicted in Fig.~\ref{fig:panel5}(a).
Chemoreceptors are organized into interacting clusters that can contain receptors with different response or adaptation properties, so that cooperative signaling and local enzyme sharing naturally couple neighboring, potentially heterogeneous, signaling units
\citep{sourjik2004functional, li2005adaptational, mello2003quantitative}.
A minimal description of the adaptation of an individual receptor is provided by a feedback loop between the activity $a \in [0,1]$ and the methylation level $m \in [0,m_0]$, as in Ref.~\cite{Lan2012}.
This feedback model has also served as a paradigmatic model for nonequilibrium signal transduction with feedback and memory \citep{Ito2015}.

To investigate the asymptotic EPR decrease in a biologically motivated heterogeneous setting, we consider three coupled adaptive signaling units, each described by a nonlinear activity and methylation model with reflecting boundaries. For $i=1,2,3$, the dynamics are given by
\begin{subequations}
    \begin{align}
        \dot{a}_i
        &=
        F_a(a_i,m_i,\nu_a)
        +
        \eta_{a,i},
        \\
        \dot{m}_i
        &=
        F_m(a_i,\nu_{m,i})
        -
        \omega_c\sum_j(m_i-m_j)
        +
        \eta_{m,i},
    \end{align}
    \label{Eq:chemotaxis}
\end{subequations}
where $F_a$ and $F_m$ provide coarse-grained descriptions of the
activity and methylation dynamics, respectively, and are defined as
\begin{subequations}
    \begin{align}
        F_a(a_i,m_i,\nu_a)
        &=
        -\nu_a
        \left[
            a_i-G(s,m_i)
        \right],
        \\
        F_m(a_i,\nu_{m,i})
        &=
        -\nu_{m,i}(a_i-a_0)
        \left[
            \beta
            -
            (1-\beta)
            \partial_m G
        \right],
    \end{align}
\end{subequations}
where $\beta$ is the chemical driving parameter. We take $\beta=1$,
corresponding to perfect adaptation \citep{Kharbanda_2024}. The function $G$ is given by
\begin{equation}
    G(s,m)
    =
    \frac{1}
    {1+s/(e^{2m})^{H}}.
\end{equation}
Here, $s$ is the ligand stimulus, $\nu_a$ and $\nu_{m,i}$ set the characteristic activity and methylation rates, respectively, with $\nu_a\gg\nu_{m,i}$, $H=1$ is the Hill coefficient, and $a_0$ is the stimulus-independent activity set point.

The terms $\eta_{a,i}$ and $\eta_{m,i}$ represent independent Gaussian white noise with strengths $2\Delta_a$ and $2\Delta_{m,i}$, respectively, where we have $\boldsymbol{\Gamma}_0=\mathbf{I}_2$. In Eq.~(\ref{Eq:chemotaxis}), interactions between neighboring signaling units are coarse-grained phenomenologically as diffusive coupling through the methylation variables, motivated by theoretical and experimental studies of cooperative receptor interactions and local sharing of adaptation enzymes within chemoreceptor clusters \citep{Endres_2006,Wingreen_PRL,Parkinson_2016,Haselwandter2014}, as depicted in Fig.~\ref{fig:panel5}(b).

We first consider equal effective temperatures for the methylation variables, that is, $\Delta_{m,1}=\Delta_{m,2}=\Delta_{m,3}\equiv\Delta_m=0.2$. Figure~\ref{fig:panel5}(c) compares identical units with $\nu_{m,i}=\nu_m=5$ to heterogeneous units with $(\nu_{m,1},\nu_{m,2},\nu_{m,3})=(5,7.5,10)$. In both cases, the EPR decreases with coupling, showing that the decreasing trend can persist in a nonlinear system with heterogeneity in the methylation rates.

To make the dependence on parameter heterogeneity explicit, we introduce a parameter $t\in[0,1]$ such that
\begin{equation}
    \nu_{m,i}(t) = \nu_m+t\delta_i,
\end{equation}
where $\delta_i$ denotes the deviation of the methylation rate of system $i$ from the identical-system value $\nu_m$. Thus, $t=0$ recovers the identical-system case $\nu_{m,i}=\nu_m$, whereas $t=1$ gives the heterogeneous parameter set.

Since the heterogeneity enters only through the methylation rates, the deterministic fixed point is unchanged along this interpolation. For $\beta=1$, it is given by $a_i^*=a_0$ and $m_i^*=m^*=\frac{1}{2}\ln\left(\frac{s a_0}{1-a_0}\right)$ for all $i$, independently of both $t$ and $\omega_c$. Linearizing about this common fixed point gives
\begin{equation}
    \mathbf M(\omega_c,t)
    =
    \begin{pmatrix}
        -\nu_a\mathbf I_3 & g\mathbf I_3\\
        -\boldsymbol \nu(t) & -\omega_c\mathbf L
    \end{pmatrix},
    \label{eq:chemotaxis-linearized}
\end{equation}
where $\boldsymbol \nu(t)=\operatorname{diag}\left(\nu_{m,1}(t),\nu_{m,2}(t),\nu_{m,3}(t)\right)$, $\mathbf I_3$ is the identity matrix, $g\equiv2\nu_a a_0(1-a_0)$, and $\mathbf L$ is the Laplacian topology matrix corresponding to the diffusive coupling between the three signaling units. For the three-unit all-to-all coupling, $\mathbf L$ has eigenvalues $(0,3,3)$. Applying the corresponding orthogonal transformation to both the activity and methylation variables gives $\mathbf L\mapsto\operatorname{diag}(0,3,3)$ and $\boldsymbol{\nu}(t)\mapsto\tilde{\boldsymbol{\nu}}(t)$. The two relative methylation coordinates are the coupled DoF, whose fluctuations are suppressed as $\omega_c$ increases, while the collective methylation coordinate and the three transformed activity coordinates constitute the residual DoF. For the linearized chemotaxis model, the residual dynamics can then be obtained explicitly. According to Eqs.~(\ref{eq:heterogeneous-transformed}) and~(\ref{eq:heterogeneous-residual}), the residual matrix reads
\begin{equation}
    \mathbf{R}_{\rm chem}(t)
    =
    \begin{pmatrix}
        -\nu_a & 0 & 0 & g \\
        0 & -\nu_a & 0 & 0 \\
        0 & 0 & -\nu_a & 0 \\
        -\tilde{\nu}_{11}(t)
        & -\tilde{\nu}_{12}(t)
        & -\tilde{\nu}_{13}(t)
        & 0
    \end{pmatrix},
    \label{eq:Rchem}
\end{equation}
where
\begin{subequations}
    \begin{align}
        \tilde{\nu}_{11}(t)
        &=
        \frac{
            \nu_{m,1}(t)
            +\nu_{m,2}(t)
            +\nu_{m,3}(t)
        }{3},
        \\
        \tilde{\nu}_{12}(t)
        &=
        \frac{
            \nu_{m,1}(t)
            -\nu_{m,2}(t)
        }{\sqrt{6}},
        \\
        \tilde{\nu}_{13}(t)
        &=
        \frac{
            \nu_{m,1}(t)
            +\nu_{m,2}(t)
            -2\nu_{m,3}(t)
        }{3\sqrt{2}}.
    \end{align}
    \label{eq:omega-tilde-elements}
\end{subequations}
As a result, the residual EPR is
\begin{equation}
\begin{split}
    \Phi_\mathrm{res}(t)
    &=
    \frac{
        \left[
            \Delta_a(\nu_m+t\bar\delta)
            +
            g\Delta_m
        \right]^2
    }{
        \nu_a\Delta_m\Delta_a
    }
    \\
    &\qquad
    +
    \frac{t^2\sigma_\delta^2}{\nu_a}
    \left[
        \frac{\Delta_a}{\Delta_m}
        +
        \frac{
            g^2
        }{
            2\nu_a^2
            +
            g(\nu_m+t\bar\delta)
        }
    \right],
\end{split}
\label{eq:phi_res_chemotaxis}
\end{equation}
where $\bar\delta=\frac{1}{3}\sum_i\delta_i$ and $\sigma_\delta^2=\frac{1}{3}\sum_i(\delta_i-\bar\delta)^2$. Equation~(\ref{eq:phi_res_chemotaxis}) makes the continuity of $\Phi_\mathrm{res}(t)$ with respect to $t$ explicit.

Additionally, the normalized residual EPR reads
\begin{equation}
\begin{split}
    \frac{\Phi_{\rm res}(t)}{\Phi_3(0,t)}
    &=
    \frac{1}{3}
    \left[
        1+
        \frac{
            \Delta_a\Delta_m g^2
            t^2\sigma_\delta^2
        }{
            \left[
                2\nu_a^2
                +
                g\left(\nu_m+t\bar{\delta}\right)
            \right]
        }
        \right.
        \\
        &\left.
        {}\times
        \frac{1}{
            \left[
                \Delta_a(\nu_m+t\bar{\delta})
                +
                g\Delta_m
            \right]^2
            +
            \Delta_a^2t^2\sigma_\delta^2
        }
    \right]
    \\
    &= \frac{1}{3}+\mathcal{O}(t^2).
\end{split}
\label{eq:normalized-residual-chemotaxis}
\end{equation}
Thus, for the heterogeneity considered here, the leading changes in the uncoupled and residual EPR cancel upon normalization, so that the strong-coupling ratio differs from its identical-system value only at $\mathcal{O}(t^2)$. Equation~(\ref{eq:normalized-residual-chemotaxis}) provides an explicit example of the local robustness to weak heterogeneity discussed above, although the leading order of the correction may differ for other forms of parameter heterogeneity.

For the parameters used in Fig.~\ref{fig:panel5}, $\frac{\Phi_{\rm res}(1)}{\Phi_3(0,1)}\simeq0.3335$, compared with $\frac{\Phi_{\rm res}(0)}{\Phi_3(0,0)}=\frac{1}{3}$ for the identical system. Accordingly, the equal-temperature curves in Fig.~\ref{fig:panel5}(c) are consistent with an approach toward nearly the same normalized strong-coupling value, despite $\Phi_3(\omega_c,0)\neq\Phi_3(\omega_c,1)$ in general.

Finally, changing only the methylation noise strength of the third unit, $\Delta_{m,3}=1.5\Delta_{m,1}$, introduces a mismatch in the effective temperatures and changes the strong-coupling behavior. For both the identical and heterogeneous parameter sets, the EPR then grows linearly for $\omega_c\gg1$, consistent with the $\mathcal{O}(\omega_c)$ strong-coupling behavior in Eq.~(\ref{eq:temperature-EPR}).


\section{Discussion}

We have characterized how coupling changes dissipation in strongly coupled linear Langevin systems and examined the corresponding results in nonlinear biophysical models. The central effect of the couplings studied here is the suppression of irreversible currents in directly coupled DoF. For identical two-dimensional subsystems, this suppression is monotonic in every nonzero coupling mode and guarantees a monotonic decrease of the corresponding EPR contribution. In more than two dimensions, the residual DoF can retain irreversible circulation. The residual EPR determines whether the strongly coupled system dissipates less than the uncoupled system. As the coupling parameter tends to infinity, the slope of the EPR is governed by the first finite-coupling correction, whose sign depends on the alignment between the residual current and the current correction induced by the coupled DoF. {We further show that when directly coupled DoF have different effective temperatures, their fluctuations vanish in the strong-coupling limit, while the total EPR can grow without bound rather than converge to the finite residual EPR.} We demonstrate EPR reduction with increasing coupling in well-studied models of bacterial chemoreceptor signaling and vertebrate hair-cell bundles.

{When the coupling admits a collective zero mode, 
the collective and relative DoF decouple, so that the coupling only affects the relative DoF.} This property is explicit in the linear Langevin systems considered here, although the collective dynamics may still be modified {by coupling} indirectly through interactions with the relative DoF, particularly in heterogeneous systems. Our biological examples illustrate this scenario: the reduced nonlinear hair-cell dynamics retain a collective coordinate (see Eq.~(\ref{eq:hair-cell-residual})), while our previous work on bacterial chemotaxis showed that coupling can reduce EPR without impairing adaptation performance~\cite{Kharbanda_2024}. Other studies have also demonstrated that coupling can preserve or enhance functional responses in hair-cell bundles and bacterial chemoreceptor networks~\cite{Barral2010,Frank2016, Dierkes2008Enhancement}. Together, these findings suggest that coupling-induced dissipation reduction need not compromise collective function, although this is not guaranteed in general, particularly for nonlinear systems.

Our results suggest a design criterion for reducing dissipation through strong coupling. Provided that the coupled DoF are maintained at the same effective temperature, coupling configurations should be chosen such that the surviving residual dynamics has a lower EPR than the uncoupled system. Particularly favorable cases arise when the residual dynamics is effectively one-dimensional or satisfies detailed balance, for which the residual EPR vanishes. Together with recent results in Refs.~\citep{Nagayama2026,Troude2026}, this criterion could support the identification of energetically favorable coupling configurations. Furthermore, the residual EPR and leading strong-coupling coefficient can be calculated from reduced equations without repeatedly solving the full Lyapunov equation, providing an advantage in computational efficiency.

The present work also points to several natural extensions. One is to develop more general criteria for characterizing the global dependence of the EPR on coupling strength in higher-dimensional linear Langevin systems, beyond the strong-coupling regime considered here. Another is to understand how coupling reshapes the EPR in nonlinear systems more generally, particularly in dynamical regimes beyond stable fixed points, including stochastic limit cycles and other oscillatory states, which would be especially relevant for biological systems such as active hair-cell bundles. Finally, extending the analysis to discrete-state nonequilibrium systems, such as coupled Markov jump processes, could help determine which features of the EPR response to coupling are specific to the present Langevin setting and which persist more generally. Together with our results, these extensions could provide a framework for identifying and designing energy-efficient coupled stochastic systems.

Beyond the presented biological systems, the question of efficiency may also be relevant for the design
of synthetic stochastic systems
\citep{Filliger2007BrownianGyrator}, including stochastic processors, noisy
computational devices, and probabilistic or thermodynamic computing architectures
\citep{Melanson2025ThermodynamicComputing,
Singh2024CMOSStochasticNanomagnets}. Continuous-variable thermodynamic
processors based on coupled OU dynamics are particularly relevant, with applications
to Gaussian sampling and linear algebra
\citep{Aifer2024ThermodynamicLinearAlgebra,
Duffield2025ThermodynamicMatrixExponentials}. In such systems,
identifying which variables to couple and the residual dynamics that remains
at strong coupling provides a direct route for assessing whether collective
operation can reduce EPR.

\textit{Acknowledgments --} This project has received funding from the European Research Council (ERC) under the European Union’s Horizon 2020 research and innovation programme (BacForce, G.A. No. 852585). Several derivations and numerical calculations were done with the assistance of LLMs such as ChatGPT.

\appendix
\section{Numerical estimation of the EPR in nonlinear systems}
\label{AppendixTrajEPR}

For the nonlinear models considered in the paper, we estimate the
steady-state EPR from simulated trajectories of the nonlinear Langevin
equation
\begin{equation}
    \frac{\mathrm{d}\mathbf{x}(t)}{\mathrm{d}t}
    =
    \mathbf{F}(\mathbf{x})
    +
    \sqrt{2\mathbf{D}}\,
    \boldsymbol{\eta}(t),
\end{equation}
where $\mathbf{F}$ is the nonlinear drift and $\mathbf{D}$ is a constant
positive-definite diffusion matrix.

Following the stochastic-energetics description of
Ref.~\cite{Sekimoto1998LangevinThermodynamics}, the entropy exchanged with the noise channels along
a trajectory is
\begin{equation}
    \mathrm{d}S_{\mathrm{ex}}
    =
    -
    \mathbf{F}^{\top}(\mathbf{x})
    \mathbf{D}^{-1}
    \circ\mathrm{d}\mathbf{x},
\end{equation}
where $\circ$ denotes the Stratonovich interpretation. In the steady state,
the average rate of change of the system entropy vanishes. The EPR is
therefore the negative of the average entropy-exchange rate,
\begin{equation}
    \Phi
    =
    -\left\langle
        \frac{\mathrm{d}S_{\mathrm{ex}}}{\mathrm{d}t}
    \right\rangle_{\mathrm{ss}}
    =
    \left\langle
        \mathbf{F}^{\top}(\mathbf{x})
        \mathbf{D}^{-1}
        \circ
        \frac{\mathrm{d}\mathbf{x}}{\mathrm{d}t}
    \right\rangle_{\mathrm{ss}}.
\label{eq:trajectory-EPR}
\end{equation}
For the linear drift
$\mathbf{F}(\mathbf{x})=\mathbf{M}\mathbf{x}$, this expression is
equivalent to Eq.~(\ref{eq:EPR_Q}).

We generate the stochastic trajectories using the Euler--Maruyama method, with integration time steps $\Delta t=5\times10^{-4}$ for the chemotaxis model and $\Delta t=5\times10^{-5}$ for the hair-cell model. After discarding an initial transient, the Stratonovich integral in Eq.~(\ref{eq:trajectory-EPR}) is evaluated using the midpoint discretization. The EPR is estimated by averaging over $10^4$ independent trajectories for each parameter set. For the chemotaxis model, reflecting boundary conditions are imposed at $a_i=0,1$ and $m_i=0,m_0$, following Ref.~\cite{Kharbanda_2024}.

To assess numerical convergence, we repeated representative simulations with progressively smaller integration timesteps while keeping the total simulated time fixed. The resulting EPR estimates showed only small quantitative variations at the finer timesteps, while the coupling-dependent trends remained unchanged.

\section{Modal decomposition}
\label{app:modal-decomposition}

We derive the modal decomposition used in Sec.~\ref{subsection:modal_decomp}.
For $n$ identical copies of the uncoupled system,
$\boldsymbol{\Gamma}=\mathbf{I}_n\otimes\boldsymbol{\Gamma}_0$,
$\boldsymbol{\Lambda}=\mathbf{I}_n\otimes\boldsymbol{\Lambda}_0$, and
$\mathbf{D}=\mathbf{I}_n\otimes\mathbf{D}_0$. The coupled dynamics and
coupling matrix are given by Eqs.~(\ref{eq:coupled-dynamics}) and
(\ref{generic_coupling_matrix}). The selector for coordinate $l$ can be
written as
\begin{equation}
    \mathbf{G}_l
    =
    \operatorname{diag}(0,\ldots,0,1,0,\ldots,0),
\end{equation}
with the nonzero entry in position $l$.

We assume that the topology matrices commute,
\begin{equation}
    [\mathbf{K}_l,\mathbf{K}_{l'}]
    =
    \mathbf{0},
    \qquad
    l,l'\in\mathcal{C}.
\end{equation}
Since the matrices $\mathbf{K}_l$ are symmetric, there exists an orthogonal
matrix $\mathbf{U}$ that diagonalizes them simultaneously,
\begin{equation}
    \mathbf{U}^{\top}\mathbf{K}_l\mathbf{U}
    =
    \operatorname{diag}
    \left(
        \lambda_{l,1},\ldots,\lambda_{l,n}
    \right).
\end{equation}
Introducing the transformation
$\mathbf{T}=\mathbf{U}\otimes\mathbf{I}_d$, the uncoupled drift and diffusion
matrices remain unchanged, while the coupling matrix becomes block diagonal.
Using $(\mathbf{P}\otimes\mathbf{Q})(\mathbf{R}\otimes\mathbf{S})
=(\mathbf{PR})\otimes(\mathbf{QS})$, we obtain
\begin{equation}
\begin{split}
    \mathbf M
    &=
    \mathbf T^\top
    \left[
    \mathbf I_n\otimes
    (\boldsymbol\Gamma_0^{-1}\boldsymbol\Lambda_0)
    -
    \omega_c\sum_{l\in\mathcal C}
    \mathbf K_l\otimes
    (\boldsymbol\Gamma_0^{-1}\mathbf G_l)
    \right]\mathbf T
    \\
    &=
    \mathbf I_n\otimes
    (\boldsymbol\Gamma_0^{-1}\boldsymbol\Lambda_0)
    -
    \omega_c\sum_{l\in\mathcal C}
    (\mathbf U^\top\mathbf K_l\mathbf U)
    \otimes
    (\boldsymbol\Gamma_0^{-1}\mathbf G_l)
    \\
    &=
    \bigoplus_{m=1}^{n}
    \boldsymbol\Gamma_0^{-1}
    \left(
    \boldsymbol\Lambda_0
    +
    \omega_c\boldsymbol\kappa_m
    \right)
\end{split}
\end{equation}

where
\begin{equation}
    \boldsymbol{\kappa}_m
    =
    -\sum_{l\in\mathcal{C}}
    \lambda_{l,m}\mathbf{G}_l.
\end{equation}
Thus, the full $nd$-dimensional system decomposes into $n$ independent
$d$-dimensional modes.

The stationary covariance is correspondingly block diagonal,
${\mathbf{S}}=\bigoplus_{m=1}^{n}\mathbf{S}_m$, where each modal
covariance satisfies
\begin{equation}
    \mathbf{M}_m\mathbf{S}_m
    +
    \mathbf{S}_m\mathbf{M}_m^{\top}
    =
    -2\mathbf{D}_0.
    \label{eq:modal-lyapunov-general}
\end{equation}
Defining $\mathbf{Q}_m=\mathbf{M}_m\mathbf{S}_m+\mathbf{D}_0$,
Eq.~(\ref{eq:EPR_Q}) gives the modal contribution
\begin{equation}
    \tilde{\Phi}_m
    =
    \operatorname{tr}
    \left[
        \mathbf{D}_0^{-1}
        \mathbf{Q}_m
        \mathbf{S}_m^{-1}
        \mathbf{Q}_m^{\top}
    \right].
    \label{eq:EPR_per_mode}
\end{equation}
Summing these independent contributions recovers Eq.~(\ref{eq:modal-epr-sum}).

\section{$d = 2$}
\label{app:d2}
 
\subsection{Modal current suppression in the strong-coupling limit}
\label{app:d2-current-suppression}

The $yy$ element of the modal covariance matrix $\mathbf{S}_m$ is
\begin{equation}
\begin{split}
    [\mathbf{S}_m]_{yy}
    &=
    \frac{
        c^2D_x
        +
        \bigl[
            bc+a(a+d+\lambda_m\omega_c)
        \bigr]D_y
    }
    {
        (a+d+\lambda_m\omega_c)
        \bigl[
            bc+a(d+\lambda_m\omega_c)
        \bigr]
    }
    \\
    &=
    \mathcal{O}(\omega_c^{-1}).
\end{split}
\end{equation}
For every nonzero mode with $\lambda_m>0$, this quantity decreases
monotonically with the coupling strength $\omega_c$. Thus, the nonzero modes
become increasingly confined in the $y_m$ direction.

Under the modal decomposition, the antisymmetric matrix $\mathbf{Q}$ also
decomposes into modal blocks of the form
\begin{equation}
    \mathbf{Q}_m
    =
    \begin{pmatrix}
        0 & -q_m \\
        q_m & 0
    \end{pmatrix},
    \qquad
    q_m
    =
    \frac{cD_x+bD_y}
    {a+d+\lambda_m\omega_c}.
\end{equation}
Therefore, for every $\lambda_m>0$, we have
$\mathbf{Q}_m\to\mathbf{0}$ as $\omega_c\to\infty$ and hence, from
Eq.~(\ref{eq:EPR_per_mode}), $\tilde{\Phi}_m\to0$. However, the local
velocity matrix
\begin{equation}
    \mathbf{v}_m
    =
    \mathbf{Q}_m\mathbf{S}_m^{-1},
    \label{eq:local_velocity}
\end{equation}
need not vanish because the covariance matrix $\mathbf{S}_m$ becomes
singular in the same limit. To quantify current suppression, we therefore
evaluate the modal current norm before taking the strong-coupling limit.

For the $m$-th mode, the stationary distribution is Gaussian,
\begin{equation}
    \rho_m(\mathbf{z}_m)
    =
    \mathcal{N}(0,\mathbf{S}_m),
    \qquad
    \mathbf{z}_m=(x_m,y_m)^\top.
\end{equation}
The corresponding modal probability current is
\begin{equation}
    \mathbf{j}_m(\mathbf{z}_m)
    =
    \rho_m(\mathbf{z}_m)\mathbf{v}_m\mathbf{z}_m.
\end{equation}
Its $L^2$ norm within the modal subspace is
\begin{equation}
    \|\mathbf{j}_m\|_{L^2_m}^2
    =
    \int_{\mathbb{R}^2}
    \rho_m^2(\mathbf{z}_m)
    \left|
        \mathbf{v}_m\mathbf{z}_m
    \right|^2
    \,\mathrm{d}\mathbf{z}_m.
\end{equation}
The density $\rho_m(\mathbf{z}_m)$ is the Gaussian density given in
Eq.~(\ref{eq:Gaussian_dist}), with covariance $\mathbf{S}_m$. Evaluating
the integral gives
\begin{equation}
    \|\mathbf{j}_m\|_{L^2_m}^2
    =
    \frac{1}
    {8\pi\sqrt{\det\mathbf{S}_m}}
    \operatorname{tr}
    \left[
        \mathbf{v}_m^\top
        \mathbf{v}_m
        \mathbf{S}_m
    \right].
    \label{L2-norm-current_1}
\end{equation}
Using the definition in Eq.~(\ref{eq:local_velocity}), we obtain
\begin{equation}
    \operatorname{tr}
    \left[
        \mathbf{v}_m^\top
        \mathbf{v}_m
        \mathbf{S}_m
    \right]
    =
    \operatorname{tr}
    \left[
        \mathbf{S}_m^{-1}
        \mathbf{Q}_m^\top
        \mathbf{Q}_m
    \right].
\end{equation}
Moreover, since
$\mathbf Q_m^\top\mathbf Q_m=q_m^2\mathbf I$,
Eq.~(\ref{L2-norm-current_1}) yields
\begin{equation}
    \|\mathbf{j}_m\|_{L^2_m}^2
    =
    \frac{q_m^2}
    {8\pi\sqrt{\det\mathbf{S}_m}}
    \operatorname{tr}
    \left[
        \mathbf{S}_m^{-1}
    \right].
    \label{eq:current_norm_eq_2}
\end{equation}

Defining $d_m=d+\lambda_m\omega_c$, the strong-coupling limit of every
nonzero mode, $\lambda_m>0$, corresponds to $d_m\to\infty$. In this limit,
the modal covariance elements scale as
\begin{subequations}
\begin{align}
    [\mathbf{S}_m]_{xx}
    &\to \frac{D_x}{a}, \\
    [\mathbf{S}_m]_{yy}
    &\sim \frac{D_y}{d_m}, \\
    [\mathbf{S}_m]_{xy}
    &\sim \frac{cD_x}{ad_m}.
\end{align}
\end{subequations}
Consequently, the determinant scales as
\begin{equation}
    \det\mathbf{S}_m
    \sim
    \frac{D_xD_y}{ad_m},
\end{equation}
whereas the trace of the inverse covariance scales as
\begin{equation}
    \operatorname{tr}
    \left[
        \mathbf{S}_m^{-1}
    \right]
    \sim
    \frac{d_m}{D_y}.
\end{equation}
Moreover, $q_m\sim(cD_x+bD_y)/d_m$. Therefore,
Eq.~(\ref{eq:current_norm_eq_2}) gives
\begin{equation}
    \|\mathbf{j}_m\|_{L^2_m}^2
    \sim
    \frac{
        \kappa^2\sqrt{aD_y}
    }
    {
        8\pi\sqrt{D_x}
    }
    d_m^{-1/2}
    \to0,
\end{equation}
where $\kappa\equiv b+cD_x/D_y$. Hence,
$\lim_{\omega_c\to\infty}\|\mathbf{j}_m\|_{L^2_m}^2=0$ for every
$\lambda_m>0$. Furthermore, it follows from the Cauchy-Schwarz inequality that 
\begin{equation}
\begin{split}
    \| \mathbf j_m \|^2_{L^1} &= \bigg(\int_{\mathbb R^2} |\mathbf j_m|  \mathrm d\mathbf z_m \bigg)^2\\
    &\leq 8 \pi \sqrt{\text{det} \mathbf S_m}\, \|\mathbf{j}_m\|^2_{L^2} \to 0.
\end{split}
\end{equation}

Thus, every nonzero modal current is suppressed in the strong-coupling
limit. Geometrically, the limiting distribution becomes singular and
concentrates on the subspace $y_m=0$. Although the local velocity field
need not vanish, its nonvanishing contribution is weighted by fluctuations
along the collapsing $y_m$ direction, causing the modal probability current
to vanish in $L^2_m$.

\subsection{Hair-cell model}
\label{app:hair-cell-model}

As introduced in the main text, we use a minimal hair-bundle model as a
nonlinear example of an actively driven system. Here, we provide the
dynamical equations, the fixed-point and stability analysis, and the
construction of the membrane-mediated nearest-neighbor coupling used in the
numerical simulations. The model describes the interaction between the
passive elastic elements of the stereocilia and the active adaptation
process driven by molecular motors, whose feedback can stabilize the bundle
or, for sufficiently strong activity, generate spontaneous oscillations
\cite{Dierkes2012}.

\begin{figure}
    \centering
    \includegraphics[width=\linewidth]{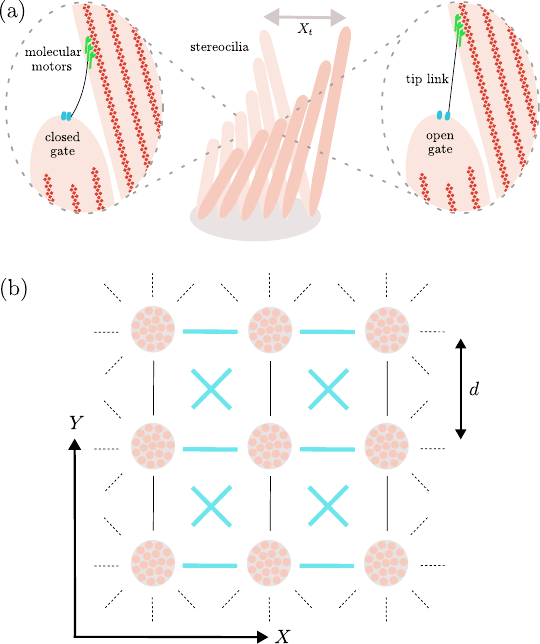}
    \caption{(a) Schematic of a single hair-cell bundle composed of rows of
stereocilia. Tip links connect neighboring stereocilia and are pulled by
molecular motors. Deflection of the bundle along its sensitive axis,
labeled $X$, changes the tip-link tension and opens or closes the
transduction channels. (b) Schematic of the square hair-bundle array
used in the simulations, showing the membrane-mediated couplings between
neighboring bundles.}
    \label{figure-hairbundle1}
\end{figure}

To describe this mechanically coupled feedback loop at the level of a
single hair bundle, we use a minimal two-variable model that tracks the
dynamics of the tip position $X\in\mathbb{R}$ and the motor coordinate
$C\in\mathbb{R}$ \cite{hair_cell_Roldan_2021},
\begin{subequations}
\begin{align}
    \lambda_x\dot{X}
    &=
    -\frac{\partial V}{\partial X}
    +
    \sqrt{2k_{\mathrm B}T\lambda_x}\,
    \eta_x(t),
    \\
    \lambda_c\dot{C}
    &=
    -\frac{\partial V}{\partial C}
    -
    F_{\mathrm{act}}(X,C)
    +
    \sqrt{2k_{\mathrm B}T_{\mathrm{eff}}\lambda_c}\,
    \eta_c(t).
\end{align}
\label{single_haircell_LE}
\end{subequations}
Here, $\lambda_x>0$ and $\lambda_c>0$ are friction coefficients, and
$\eta_x(t)$ and $\eta_c(t)$ are independent unit Gaussian white noises.
The parameter $T$ denotes the ambient temperature, whereas
$T_{\mathrm{eff}}>T$ is an effective temperature capturing motor-induced
fluctuations.

The active force generated by the molecular motors is
\begin{equation}
    F_{\mathrm{act}}(X,C)
    =
    F_{\mathrm{max}}
    \left[
        1-SP_0(X,C)
    \right],
    \label{F_act}
\end{equation}
where $F_{\mathrm{max}}$ is the maximal active force and $S$ controls the
calcium-mediated feedback. The elastic elements of the bundle and the
transduction channels are described by the conservative potential
\begin{equation}
\begin{split}
    V(X,C)
    &=
    \frac{
        k_{\mathrm{gs}}(X-C)^2
        +
        k_{\mathrm{sp}}X^2
    }{2}
    \\
    &\quad
    -
    Nk_{\mathrm B}T
    \ln\left[
        \exp\left(
            \frac{k_{\mathrm{gs}}D(X-C)}
            {Nk_{\mathrm B}T}
        \right)
        +
        A
    \right].
\end{split}
\end{equation}
Here, $k_{\mathrm{gs}}$ and $k_{\mathrm{sp}}$ are the stiffness
coefficients of the gating spring and pivot spring, respectively, and $D$
is the gating swing of a transduction channel. The dimensionless gating
factor is
\begin{equation}
    A
    =
    \exp\left[
        \frac{
            \Delta G+k_{\mathrm{gs}}D^2/(2N)
        }
        {k_{\mathrm B}T}
    \right],
\end{equation}
where $\Delta G$ is the energy difference between the open and closed
states of the channel and $N$ is the number of transduction elements.

The open probability of the transduction channels is
\begin{equation}
    P_0(X,C)
    =
    \left[
        1
        +
        A\exp\left(
            -\frac{k_{\mathrm{gs}}D(X-C)}
            {Nk_{\mathrm B}T}
        \right)
    \right]^{-1}.
\end{equation}
An increase in the tip--motor extension $X-C$ raises the open
probability $P_0$, which, through the feedback factor $S$ in
Eq.~(\ref{F_act}), reduces the active force. This negative feedback links
channel gating to motor activity and helps stabilize the hair-bundle
dynamics.

The fixed point $(X^*,C^*)$ follows from the nonlinear steady-state
conditions,
\begin{subequations}
\begin{align}
    X^*
    &=
    -\frac{F_{\mathrm{max}}}{k_{\mathrm{sp}}}
    \left(
        1-SP_0^*
    \right),
    \\
    C^*
    &=
    X^*
    -
    \left[
        DP_0^*
        +
        \frac{F_{\mathrm{max}}}{k_{\mathrm{gs}}}
        \left(
            1-SP_0^*
        \right)
    \right],
\end{align}
\label{fixedpoint_haricell}
\end{subequations}
where $P_0^*\equiv P_0(X^*,C^*)$.

A linear-stability analysis around this fixed point reveals two oscillatory stability crossings as $F_{\mathrm{max}}$ is varied. The fixed point is stable for $F_{\mathrm{max}}<F_c^-$ and $F_{\mathrm{max}}>F_c^+$, and unstable for $F_c^-<F_{\mathrm{max}}<F_c^+$. The trace of the Jacobian $\mathbf{J}^*$ evaluated at $(X^*,C^*)$ is
\begin{equation}
\begin{split}
    \operatorname{tr}(\mathbf{J}^*)
    &=
    \frac{
        -k_{\mathrm{sp}}
        -k_{\mathrm{gs}}
        +
        \gamma Dk_{\mathrm{gs}}
    }{\lambda_x}
    \\
    &\quad
    +
    \frac{
        -k_{\mathrm{gs}}
        +
        \gamma Dk_{\mathrm{gs}}
        -
        SF_{\mathrm{max}}\gamma
    }{\lambda_c}.
\end{split}
\end{equation}
The critical force is determined implicitly by
$\operatorname{tr}(\mathbf{J}^*)=0$, giving
\begin{equation}
    F_c(S)
    =
    -\frac{
        k_{\mathrm{sp}}\lambda_c
        +
        k_{\mathrm{gs}}(1-\gamma D)
        (\lambda_x+\lambda_c)
    }{
        \lambda_xS\gamma
    },
    \label{critical_force}
\end{equation}
where
\begin{equation}
    \gamma
    \equiv
    \frac{k_{\mathrm{gs}}D}{Nk_{\mathrm B}T}
    P_0^*(1-P_0^*)
    \geq0
\end{equation}
is evaluated at the critical fixed point. Equation~(\ref{critical_force})
is implicit because $P_0^*$, and therefore $\gamma$, depends on
$F_{\mathrm{max}}$ through the fixed-point conditions.

For the Jacobian considered here, a zero of $\operatorname{tr}(\mathbf{J}^*)$ with $\det(\mathbf{J}^*)>0$, accompanied by a change in sign of the trace, corresponds to an oscillatory stability crossing.
For the model, this condition is satisfied at
two critical forces, denoted by $F_c^-$ and $F_c^+$. For
$F_{\mathrm{max}}<F_c^-$, the eigenvalues have negative real parts
and the fixed point is stable. At $F_c^-$, a complex-conjugate pair of
eigenvalues crosses the imaginary axis and the fixed point loses
stability, marking the onset of the oscillatory regime. At $F_c^+$,
the real parts cross back to negative values and the stable fixed-point
regime is recovered.

The active force and the channel-opening probability form a negative
feedback loop. Increasing $F_{\mathrm{max}}$ shifts the fixed point and
increases the active input. At the same time, the resulting change in
$X^*-C^*$ raises $P_0^*$ and reduces the factor $1-SP_0^*$ in the active
force. Increasing $S$ strengthens this feedback and increases the critical
force required to destabilize the fixed point.

The motor coordinate responds more slowly than the tip position, with a characteristic relaxation time that is typically larger than that of the tip. This delay introduces the phase lag required for the oscillatory instability. As $F_{\mathrm{max}}$ increases, the competition between active forcing and the restoring feedback gives rise to the intermediate oscillatory instability described above.

\subsubsection{Nearest-neighbor coupling}
\label{app:hair-cell-coupling}

To study membrane-mediated coupling, we consider a square array of linear
size $L$ containing $n=L^2$ identical hair bundles, as in
Ref.~\cite{Dierkes2012}. The variables of the bundle at lattice site
$(i,j)$ are denoted by $X_{(i,j)}$ and $C_{(i,j)}$, where
$i,j\in\{1,\ldots,L\}$. Here, we consider only deflections along the
$X$ direction.

The overlying membrane is modeled by elastic elements connecting
neighboring bundles. Consider a connection between the bundle at $(i,j)$
and the bundle at $(i+k,j+l)$, where
$k,l\in\{-1,0,1\}$. The corresponding elastic potential is
\begin{equation}
    U^{i,j}_{k,l}
    =
    \frac{\omega_c}{2}
    \left[
        L^{i,j}_{k,l}
        -
        L^0_{k,l}
    \right]^2,
\end{equation}
where the instantaneous length is
\begin{equation}
    L^{i,j}_{k,l}
    =
    \sqrt{
        \left[
            X_{(i+k,j+l)}
            -
            X_{(i,j)}
            +
            kd
        \right]^2
        +
        l^2d^2
    },
\end{equation}
$d$ is the lattice spacing, and the rest length is
\begin{equation}
    L^0_{k,l}
    =
    d\sqrt{k^2+l^2}.
\end{equation}
The rest length corresponds to the configuration in which all bundles are
at their mechanical rest positions.

We introduce the symmetric neighbor set $\mathcal{N}=\{(\pm1,0),(0,\pm1),(\pm1,\pm1)\}.$
For bundles at the boundary of the array, the sums over $\mathcal{N}$
include only those terms for which the neighboring site $(i+k,j+l)$ lies
within the array, corresponding to open boundary conditions. Since
$\mathcal{N}$ contains both orientations of each elastic connection, the
total membrane energy is
\begin{equation}
    U_{\mathrm{mem}}
    =
    \frac{\omega_c}{4}
    \sum_{i,j=1}^{L}
    \sum_{(k,l)\in\mathcal{N}}
    \left[
        L^{i,j}_{k,l}
        -
        L^0_{k,l}
    \right]^2.
    \label{eq:hair-total-membrane-energy}
\end{equation}
Here the additional factor of  $1/2$ corrects for the double counting of each
elastic connection in the symmetric neighbor sum.

If we assume that the bundle deflections are small compared with the
lattice spacing,
\begin{equation}
    \left|
        \Delta X^{i,j}_{k,l}
    \right|
    =
    \left|
        X_{(i+k,j+l)}
        -
        X_{(i,j)}
    \right|
    \ll d,
\end{equation}
we obtain
\begin{equation}
    L^{i,j}_{k,l}
    -
    L^0_{k,l}
    =
    \frac{k}{\sqrt{k^2+l^2}}
    \Delta X^{i,j}_{k,l}
    +
    \mathcal{O}\left(
        \frac{
            (\Delta X^{i,j}_{k,l})^2
        }{d}
    \right).
\end{equation}
The corresponding elastic-energy contribution therefore becomes
\begin{equation}
    U^{i,j}_{k,l}
    \approx
    \frac{\omega_{k,l}}{2}
    \left(
        X_{(i+k,j+l)}
        -
        X_{(i,j)}
    \right)^2,
    \label{eq:hair-membrane-potential}
\end{equation}
where $\omega_{k,l}=\omega_c \frac{k^2}{k^2+l^2}.$ Thus, the linearized coupling vanishes for vertically adjacent bundles,
while $\omega_{\pm1,0}=\omega_c$ for horizontal neighbors and
$\omega_{\pm1,\pm1}=\omega_c/2$ for diagonal neighbors. Differentiating the linearized membrane energy gives the corresponding
force contribution
\begin{equation}
    \frac{\partial U_{\mathrm{mem}}}
    {\partial X_{(i,j)}}
    =
    \sum_{(k,l)\in\mathcal{N}}
    \omega_{k,l}
    \left(
        X_{(i,j)}
        -
        X_{(i+k,j+l)}
    \right).
    \label{eq:hair-membrane-force}
\end{equation}
As above, terms for which $(i+k,j+l)$ lies outside the array are omitted.

Substituting Eq.~(\ref{eq:hair-membrane-force}) into the single-bundle
dynamics gives the membrane-coupled hair-bundle model stated in Eq.~(\ref{eq:hair-cell}) of
the main text. The membrane force acts only on the bundle-position
coordinate $X_{(i,j)}$, while the motor coordinate $C_{(i,j)}$ is
coupled indirectly through the internal hair-bundle dynamics.

\subsubsection{Simulation parameters and numerical results}

For the numerical results in Fig.~\ref{fig:panel2}(b) of the main text, we use
$\lambda_x=2.8\times10^{-3}\,\mathrm{pN\,s\,nm^{-1}}$,
$\lambda_c=10^{-2}\,\mathrm{pN\,s\,nm^{-1}}$,
$k_{\mathrm{gs}}=0.75\,\mathrm{pN\,nm^{-1}}$,
$k_{\mathrm{sp}}=0.60\,\mathrm{pN\,nm^{-1}}$,
$D=61\,\mathrm{nm}$, $N=50$, $S=0.94$,
$\Delta G=41.4\,\mathrm{pN\,nm}$,
$k_{\mathrm B}T=4.143\,\mathrm{pN\,nm}$,
$T_{\mathrm{eff}}/T=1.5$, and
$F_{\mathrm{max}}=31\,\mathrm{pN}$.
Equivalently,
$k_{\mathrm B}T_{\mathrm{eff}}\simeq6.215\,\mathrm{pN\,nm}$.
For these parameters, the two stability boundaries are
$F_c^-\approx52.8\,\mathrm{pN}$ and
$F_c^+\approx85.7\,\mathrm{pN}$.
Thus, the numerical results in Fig.~\ref{fig:panel2} at
$F_{\mathrm{max}}=31\,\mathrm{pN}$ lie within the low-force stable
fixed-point regime.

To examine the influence of nonlinearities, we additionally consider
$F_{\mathrm{max}}=35$, $45$, and $52.5\,\mathrm{pN}$, keeping all other
parameters unchanged. These values remain below the lower stability
boundary $F_c^-$. For comparison, we construct a corresponding
nonlinear residual system by imposing the strong-coupling constraint
$X_{(i,j)}=\bar X$, where
$\bar X=n^{-1}\sum_{(i,j)}X_{(i,j)}$ is the collective bundle position.
The membrane forces then cancel in the collective equation, while the
individual motor coordinates remain dynamical. The resulting residual
system is
\begin{subequations}
\begin{align}
&\lambda_x\dot{\bar X}
=
-\frac{1}{n}
\sum_{(i,j)}
\frac{\partial V(\bar X,C_{(i,j)})}{\partial\bar X}
+
\sqrt{\frac{2k_{\mathrm B}T\lambda_x}{n}}\,
\eta_{\bar X},
\\
&\lambda_c\dot C_{(i,j)}
=
-\frac{\partial V(\bar X,C_{(i,j)})}{\partial C_{(i,j)}}
\nonumber\\
&\qquad \qquad 
-F_{\mathrm{act}}(\bar X,C_{(i,j)}) +\sqrt{2k_{\mathrm B}T_{\mathrm{eff}}\lambda_c}\,
\eta_c^{(i,j)},
\end{align}
\label{eq:hair-cell-residual}
\end{subequations}
where $\eta_{\bar X}$ and $\eta_c^{(i,j)}$ are independent unit
Gaussian white noises.

Unlike identical linear systems, the nonlinear dynamics do not
decouple into independent spatial modes, since the nonlinear forces
couple the collective position to the individual motor coordinates.
This resembles the case of heterogeneous systems, where the
absence of independent modal dynamics also requires the residual
behavior to be determined from the coupled system as a whole.
Consequently, the nonlinear residual EPR need not coincide with
the $1/n$ prediction of the identical linear theory.

\begin{figure}[t]
    \centering
    \includegraphics[width=\linewidth]{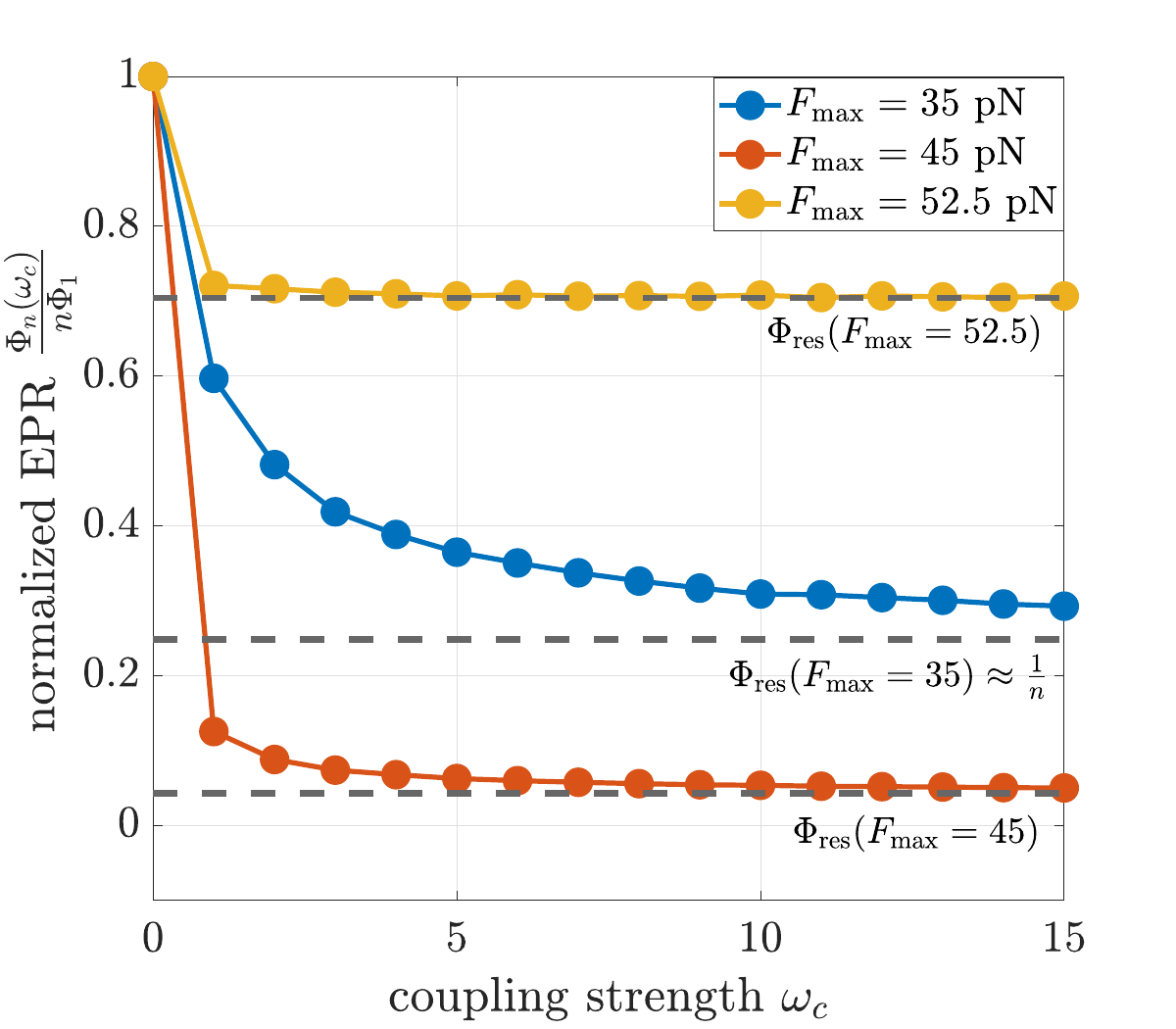}
    \caption{
    Normalized EPR as a function of coupling strength for
    $F_{\mathrm{max}}=35$, $45$, and $52.5\,\mathrm{pN}$ and $n = 2\times 2$.
    Black dashed lines indicate the corresponding residual EPRs
    obtained from the nonlinear residual system in
    Eq.~(\ref{eq:hair-cell-residual}).
    }
    \label{fig:hair-cell-nonlinear}
\end{figure}

As shown in Fig.~\ref{fig:hair-cell-nonlinear}, the normalized EPR
decreases with increasing coupling for all three active forces.
At smaller active forces, the limiting EPR remains close to the
$1/n$ prediction of the linear theory. As the active force increases,
however, deviations from this prediction become pronounced,
particularly near the Hopf bifurcation. Nevertheless, the numerical
results approach the corresponding EPRs of the nonlinear residual
system. These observations demonstrate the robustness of the
strong-coupling mechanism identified in the linear theory for
the nonlinear hair-cell model considered here, while highlighting
that nonlinearities can substantially modify the limiting
dissipation. A general extension of our analytical results to
nonlinear systems, including conditions guaranteeing EPR
reduction, remains beyond the scope of the present work.

\section{$d \geq 3$}
\label{app:high-dimensional}
\subsection{Block representation for each mode}
\label{app:uv-decomposition}

For each nonzero mode, we use the ordering
$\mathbf{x}_m=(\mathbf{u}_m^{\top},\mathbf{v}_m^{\top})^{\top}$ introduced
in the main text, where $\mathbf{u}_m\in\mathbb{R}^k$ contains the directly
coupled variables and $\mathbf{v}_m\in\mathbb{R}^{d-k}$ the residual
variables. The modal drift therefore has the block form in
Eq.~(\ref{eq:modal-block-drift-main}), with
$\mathbf{A}_0=\boldsymbol{\Gamma}_u^{-1}\boldsymbol{\kappa}_{u,m}$,
$\mathbf{A}=\boldsymbol{\Gamma}_u^{-1}\boldsymbol{\Lambda}_{uu}$,
$\mathbf{B}=\boldsymbol{\Gamma}_u^{-1}\boldsymbol{\Lambda}_{uv}$,
$\mathbf{C}=\boldsymbol{\Gamma}_v^{-1}\boldsymbol{\Lambda}_{vu}$, and
$\mathbf{R}=\boldsymbol{\Gamma}_v^{-1}\boldsymbol{\Lambda}_{vv}$.
The mode index on these blocks is suppressed below.

The diffusion matrix has the corresponding block structure
$\mathbf{D}_0=\operatorname{diag}(\mathbf{D}_u,\mathbf{D}_v)$. Using the block decomposition of the modal covariance introduced in
Eq.~(\ref{eq:covariance-block-main}), the covariance satisfies
\begin{equation}
    \mathbf{M}_m\mathbf{S}_m
    +
    \mathbf{S}_m\mathbf{M}_m^\top
    =
    -2\mathbf{D}_0.
    \label{eq:modal-lyapunov}
\end{equation}

The modal irreversible-current matrix
\begin{equation}
    \mathbf{Q}_m
    =
    \mathbf{M}_m\mathbf{S}_m+\mathbf{D}_0
    \label{eq:modal-Q}
\end{equation}
is antisymmetric. Writing its block rows as
$\mathbf{Q}_u=(\mathbf{Q}_{uu}\ \mathbf{Q}_{uv})$ and
$\mathbf{Q}_v=(\mathbf{Q}_{vu}\ \mathbf{Q}_{vv})$, Eq.~(\ref{eq:EPR_per_mode})
gives
\begin{equation}
\begin{split}
    \tilde{\Phi}_m
    &=
    \operatorname{tr}
    \left[
        \mathbf{D}_u^{-1}
        \mathbf{Q}_u
        \mathbf{S}_m^{-1}
        \mathbf{Q}_u^\top
    \right]
    +
    \operatorname{tr}
    \left[
        \mathbf{D}_v^{-1}
        \mathbf{Q}_v
        \mathbf{S}_m^{-1}
        \mathbf{Q}_v^\top
    \right]
    \\
    &=
    \tilde{\Phi}_u+\tilde{\Phi}_v.
    \label{eq:modal-epr-blocks-app}
\end{split}
\end{equation}
The first term is associated with irreversible currents in the directly
coupled variables, whereas the second is associated with currents in the
residual variables.

\subsection{Convergence of modal EPR}
\label{app:residual-convergence}

We now determine the strong-coupling behavior of a mode with nonzero
coupling eigenvalues. For a fixed mode $m$, we suppress the mode index and
use the block form in Eq.~(\ref{eq:modal-block-drift-main}). We assume that
$\mathbf{A}_0$ is diagonal and Hurwitz and that the residual drift matrix
$\mathbf{R}$ is Hurwitz.

\subsubsection{Strong-coupling expansion of the covariance}
\label{app:covariance-expansion}

Using the block forms of the drift, diffusion, and covariance matrices
introduced above, the modal Lyapunov equation
Eq.~(\ref{eq:modal-lyapunov}) gives
\begin{subequations}
\begin{align}
    &(\omega_c\mathbf{A}_0+\mathbf{A})\mathbf{S}_{uu}
    +
    \mathbf{S}_{uu}(\omega_c\mathbf{A}_0+\mathbf{A})^\top
    \notag\\
    &\qquad \qquad \qquad \qquad +\mathbf{B}\mathbf{S}_{uv}^\top
    +
    \mathbf{S}_{uv}\mathbf{B}^\top
    =
    -2\mathbf{D}_u,
    \\
    &(\omega_c\mathbf{A}_0+\mathbf{A})\mathbf{S}_{uv}
    +
    \mathbf{B}\mathbf{S}_{vv}
    +
    \mathbf{S}_{uu}\mathbf{C}^\top
    +
    \mathbf{S}_{uv}\mathbf{R}^\top
    =
    \mathbf{0},
    \\
    &\mathbf{C}\mathbf{S}_{uv}
    +
    \mathbf{R}\mathbf{S}_{vv}
    +
    \mathbf{S}_{uv}^\top\mathbf{C}^\top
    +
    \mathbf{S}_{vv}\mathbf{R}^\top
    =
    -2\mathbf{D}_v.
\end{align}
\label{eq:block-lyapunov}
\end{subequations}

Since the Lyapunov equation is linear in the entries of the covariance and
its coefficients depend linearly on $\omega_c$, the covariance elements are
rational functions of $\omega_c$. The covariance blocks therefore admit
Laurent expansions about $\omega_c \gg 1$,
\begin{subequations}
\begin{align}
    \mathbf{S}_{uu}
    &=
    \sum_{r=-\infty}^{N_u}
    \mathbf{X}_r\omega_c^r,
    \\
    \mathbf{S}_{uv}
    &=
    \sum_{r=-\infty}^{N_{uv}}
    \mathbf{Y}_r\omega_c^r,
    \\
    \mathbf{S}_{vv}
    &=
    \sum_{r=-\infty}^{N_v}
    \mathbf{Z}_r\omega_c^r,
\end{align}
\label{eq:covariance-laurent}
\end{subequations}
for some finite integers $N_u$, $N_{uv}$, and $N_v$.

Substituting Eq.~(\ref{eq:covariance-laurent}) into
Eq.~(\ref{eq:block-lyapunov}) and collecting equal powers of $\omega_c$
gives, for every integer $r$,
\begin{subequations}
\begin{align}
    &\mathbf{A}_0\mathbf{X}_{r-1}
    +
    \mathbf{X}_{r-1}\mathbf{A}_0^\top
    +
    \mathbf{A}\mathbf{X}_r
    +
    \mathbf{X}_r\mathbf{A}^\top
    \notag\\
    &\qquad \qquad\qquad  
    +
    \mathbf{B}\mathbf{Y}_r^\top
    +
    \mathbf{Y}_r\mathbf{B}^\top
    =
    -2\mathbf{D}_u\delta_{r,0},
    \\
    &\mathbf{A}_0\mathbf{Y}_{r-1}
    +
    \mathbf{A}\mathbf{Y}_r
    +
    \mathbf{B}\mathbf{Z}_r
    +
    \mathbf{X}_r\mathbf{C}^\top
    +
    \mathbf{Y}_r\mathbf{R}^\top
    =
    \mathbf{0},
    \\
    &\mathbf{C}\mathbf{Y}_r
    +
    \mathbf{R}\mathbf{Z}_r
    +
    \mathbf{Y}_r^\top\mathbf{C}^\top
    +
    \mathbf{Z}_r\mathbf{R}^\top
    =
    -2\mathbf{D}_v\delta_{r,0}.
\end{align}
\label{eq:covariance-coefficients}
\end{subequations}

We first show that no covariance block contains a positive power of
$\omega_c$. Let
$P=\max(N_u,N_{uv},N_v)$ and suppose that $P\geq1$. Setting $r=P+1$ in
the first two equations of Eq.~(\ref{eq:covariance-coefficients}) gives
\begin{subequations}
\begin{align}
    \mathbf{A}_0\mathbf{X}_P
    +
    \mathbf{X}_P\mathbf{A}_0^\top
    &=
    \mathbf{0},
    \\
    \mathbf{A}_0\mathbf{Y}_P
    &=
    \mathbf{0}.
\end{align}
\end{subequations}
Since $\mathbf{A}_0$ is Hurwitz, the first equation has the unique solution
$\mathbf{X}_P=\mathbf{0}$. Moreover, $\mathbf{A}_0$ is invertible, so
$\mathbf{Y}_P=\mathbf{0}$. Setting $r=P$ in the third equation of
Eq.~(\ref{eq:covariance-coefficients}) then gives
\begin{equation}
    \mathbf{R}\mathbf{Z}_P
    +
    \mathbf{Z}_P\mathbf{R}^\top
    =
    \mathbf{0}.
\end{equation}
Since $\mathbf{R}$ is Hurwitz, this equation has the unique solution
$\mathbf{Z}_P=\mathbf{0}$. Repeating the same argument successively for
the remaining positive powers yields
\begin{equation}
    \mathbf{X}_r
    =
    \mathbf{Y}_r
    =
    \mathbf{Z}_r
    =
    \mathbf{0},
    \qquad
    r\geq1.
\end{equation}

We next consider the constant terms. Setting $r=1$ in the first two
equations of Eq.~(\ref{eq:covariance-coefficients}), after the positive
powers have been eliminated, gives
\begin{subequations}
\begin{align}
    \mathbf{A}_0\mathbf{X}_0
    +
    \mathbf{X}_0\mathbf{A}_0^\top
    &=
    \mathbf{0},
    \\
    \mathbf{A}_0\mathbf{Y}_0
    &=
    \mathbf{0}.
\end{align}
\end{subequations}
It follows that
$\mathbf{X}_0=\mathbf{0}$ and $\mathbf{Y}_0=\mathbf{0}$. By contrast, the
residual covariance retains a finite constant term. Setting $r=0$ in the
third equation of Eq.~(\ref{eq:covariance-coefficients}) gives
\begin{equation}
    \mathbf{R}\mathbf{Z}_0
    +
    \mathbf{Z}_0\mathbf{R}^\top
    =
    -2\mathbf{D}_v.
\label{eq:residual-lyapunov}
\end{equation}
Thus, $\mathbf{Z}_0$ is the stationary covariance of the residual dynamics
with drift $\mathbf{R}$ and diffusion $\mathbf{D}_v$.

The leading inverse-coupling coefficients follow by setting $r=0$ in the
first two equations of Eq.~(\ref{eq:covariance-coefficients}),
\begin{subequations}
\begin{align}
    \mathbf{A}_0\mathbf{X}_{-1}
    +
    \mathbf{X}_{-1}\mathbf{A}_0^\top
    &=
    -2\mathbf{D}_u,
    \\
    \mathbf{A}_0\mathbf{Y}_{-1}
    +
    \mathbf{B}\mathbf{Z}_0
    &=
    \mathbf{0}.
\end{align}
\label{eq:leading-covariance-coefficients}
\end{subequations}
Since $\mathbf A_0$ and $\mathbf R$ are Hurwitz and
$\mathbf D_u,\mathbf D_v\succ0$, the corresponding Lyapunov equations
have unique positive-definite solutions $\mathbf X_{-1}$ and
$\mathbf Z_0$, respectively.

For notational convenience, we define
$\mathbf{X}_1\equiv\mathbf{X}_{-1}$,
$\mathbf{Y}_1\equiv\mathbf{Y}_{-1}$, and
$\mathbf{Z}_1\equiv\mathbf{Z}_{-1}$. The covariance blocks therefore have
the strong-coupling expansions
\begin{subequations}
\begin{align}
    \mathbf{S}_{uu}
    &=
    \frac{\mathbf{X}_1}{\omega_c}
    +
    \mathcal{O}(\omega_c^{-2}),
    \\
    \mathbf{S}_{uv}
    &=
    \frac{\mathbf{Y}_1}{\omega_c}
    +
    \mathcal{O}(\omega_c^{-2}),
    \\
    \mathbf{S}_{vv}
    &=
    \mathbf{Z}_0
    +
    \frac{\mathbf{Z}_1}{\omega_c}
    +
    \mathcal{O}(\omega_c^{-2}).
\end{align}
\label{eq:covariance-strong-coupling}
\end{subequations}
Hence, the fluctuations of the directly coupled variables and their
correlations with the residual variables vanish as $\omega_c^{-1}$, while
the residual covariance approaches the finite matrix $\mathbf{Z}_0$.

Notably, the convergence in Eq.~(\ref{eq:covariance-strong-coupling}) does not require $\mathbf D_u$ or $\mathbf D_v$ to be diagonal. Moreover, it persists even when the diffusion matrix has off-diagonal blocks between the coupled and residual DoF. Such terms modify the leading covariance correction $\mathbf Y_1$, but the limiting residual covariance $\mathbf Z_0$ remains determined by Eq.~(\ref{eq:residual-lyapunov}).

\subsubsection{Suppression of currents of coupled DoF and their EPR contribution}
\label{app:coupled-suppression}

From Eq.~(\ref{eq:leading-covariance-coefficients}),
\begin{equation}
    \mathbf{A}_0\mathbf{X}_1
    +
    \mathbf{X}_1\mathbf{A}_0^\top
    =
    -2\mathbf{D}_u.
    \label{eq:X1-lyapunov}
\end{equation}
Since $\mathbf{A}_0$ and $\mathbf{D}_u$ are diagonal, this equation also
shows that $\mathbf{X}_1$ is diagonal. Therefore, we have that
\begin{equation}
    \mathbf{A}_0\mathbf{X}_1
    =
    \mathbf{X}_1\mathbf{A}_0^\top
    =
    -\mathbf{D}_u.
    \label{eq:A0X1-cancellation}
\end{equation}
This cancellation is essential for the suppression of the coupled-variable
EPR contribution. Its breakdown for different effective temperatures is
discussed in Appendix~\ref{app:effective-temperatures}.

Using the covariance expansion in
Eq.~(\ref{eq:covariance-strong-coupling}), the blocks in the coupled row of
the irreversible-current matrix satisfy
\begin{align}
    \mathbf{Q}_{uu}
    &=
    \mathbf{A}_0\mathbf{X}_1
    +
    \mathbf{D}_u
    +
    \mathcal{O}(\omega_c^{-1})
    =
    \mathcal{O}(\omega_c^{-1}),
    \label{eq:Quu-scaling}
    \\
    \mathbf{Q}_{uv}
    &=
    \mathbf{A}_0\mathbf{Y}_1
    +
    \mathbf{B}\mathbf{Z}_0
    +
    \mathcal{O}(\omega_c^{-1})
    =
    \mathcal{O}(\omega_c^{-1}).
    \label{eq:Quv-scaling}
\end{align}
The leading term in Eq.~(\ref{eq:Quu-scaling}) vanishes by
Eq.~(\ref{eq:A0X1-cancellation}), while that in
Eq.~(\ref{eq:Quv-scaling}) vanishes by the second equation in
Eq.~(\ref{eq:leading-covariance-coefficients}). Thus,
\begin{equation}
    \mathbf{Q}_u
    =
    \mathcal{O}(\omega_c^{-1}).
    \label{eq:Qu-scaling}
\end{equation}

The block-inversion formula applied to
Eq.~(\ref{eq:covariance-strong-coupling}) gives
\begin{equation}
    \mathbf{S}_m^{-1}
    =
    \begin{pmatrix}
        \mathcal{O}(\omega_c) & \mathcal{O}(1) \\
        \mathcal{O}(1) &
        \mathbf{Z}_0^{-1}
        +
        \mathcal{O}(\omega_c^{-1})
    \end{pmatrix}.
    \label{eq:inverse-covariance-scaling}
\end{equation}
The current associated with the coupled DoF is
$\mathbf{j}_u(\mathbf{x}_m)
=
p_m(\mathbf{x}_m)\mathbf{Q}_u\mathbf{S}_m^{-1}\mathbf{x}_m$,
where $\mathbf{Q}_u=(\mathbf{Q}_{uu},\mathbf{Q}_{uv})$. Its $L^1$ norm is therefore
\begin{equation}
    \|\mathbf{j}_u\|_{L^1}
    =
    \int
    p_m(\mathbf{x}_m)
    \left\|
        \mathbf{Q}_u\mathbf{S}_m^{-1}\mathbf{x}_m
    \right\|_2
    \,d\mathbf{x}_m.
\end{equation}
Applying the Cauchy--Schwarz inequality gives
\begin{equation}
    \begin{split}
    \|\mathbf{j}_u\|_{L^1}^2
    &\leq
    \left(\int p_m(\mathbf{x}_m)\,d\mathbf{x}_m\right) \\
    &\qquad \times
    \left[
    \int
    p_m(\mathbf{x}_m)
    \left\|
        \mathbf{Q}_u\mathbf{S}_m^{-1}\mathbf{x}_m
    \right\|_2^2
    \,d\mathbf{x}_m
    \right] \\
    &=
    \int
    p_m(\mathbf{x}_m)\,
    \mathbf{x}_m^\top
    \mathbf{S}_m^{-1}
    \mathbf{Q}_u^\top\mathbf{Q}_u
    \mathbf{S}_m^{-1}
    \mathbf{x}_m
    \,d\mathbf{x}_m \\
    &=
    \operatorname{tr}
    \left[
        \mathbf{S}_m^{-1}
        \mathbf{Q}_u^\top\mathbf{Q}_u
        \mathbf{S}_m^{-1}
        \mathbf{S}_m
    \right] \\
    &=
    \operatorname{tr}
    \left[
        \mathbf{Q}_u
        \mathbf{S}_m^{-1}
        \mathbf{Q}_u^\top
    \right] \to 0.
    \end{split}
    \label{eq:ju-L1-bound-app}
\end{equation}
where we used that
$\int p_m(\mathbf{x}_m)\mathbf{x}_m\mathbf{x}_m^\top d\mathbf{x}_m=\mathbf{S}_m$. The convergence in Eq.~(\ref{eq:ju-L1-bound}) then follows from Eqs~(\ref{eq:Qu-scaling}) and (\ref{eq:inverse-covariance-scaling}).

Furthermore, using the modal EPR decomposition derived above,
\begin{equation}
    \tilde{\Phi}_u
    =
    \operatorname{tr}
    \left[
        \mathbf{D}_u^{-1}
        \mathbf{Q}_u
        \mathbf{S}_m^{-1}
        \mathbf{Q}_u^\top
    \right].
\end{equation}
Both blocks of $\mathbf{Q}_u$ are of order $\omega_c^{-1}$, whereas the
largest block of $\mathbf{S}_m^{-1}$ is of order $\omega_c$. Consequently,
\begin{equation}
    \tilde{\Phi}_u
    =
    \mathcal{O}(\omega_c^{-1})
    \longrightarrow 0
    \qquad
    \text{as }
    \omega_c\to\infty.
    \label{eq:phi-u-limit}
\end{equation}
Thus, strong coupling suppresses the EPR contribution associated with the
directly coupled variables (see Fig.~\ref{fig:strong-coupling-scalings}(b)).

\subsubsection{Faster suppression for a single coupled variable}
\label{app:faster-suppression}

The decay of the coupled contribution is faster when
\begin{equation}
    \mathbf{A}\mathbf{D}_u-\mathbf{D}_u\mathbf{A}^{\top}=\mathbf{0}.
    \label{eq:fast-detailed-balance}
\end{equation}
Since $\mathbf A_0$ and $\mathbf D_u$ are diagonal, Eq.~(\ref{eq:fast-detailed-balance}) implies $(\omega_c\mathbf A_0+\mathbf A)\mathbf D_u=\mathbf D_u(\omega_c\mathbf A_0+\mathbf A)^\top$. Thus, the dynamics within the directly coupled variables satisfies the linear detailed-balance condition, so the leading internal irreversible-current contribution vanishes. The coupled EPR contribution is then determined by the mixed current and decays as $\tilde{\Phi}_u=\mathcal O(\omega_c^{-2})$.

\begin{figure}
    \centering
    \includegraphics[width=\linewidth]{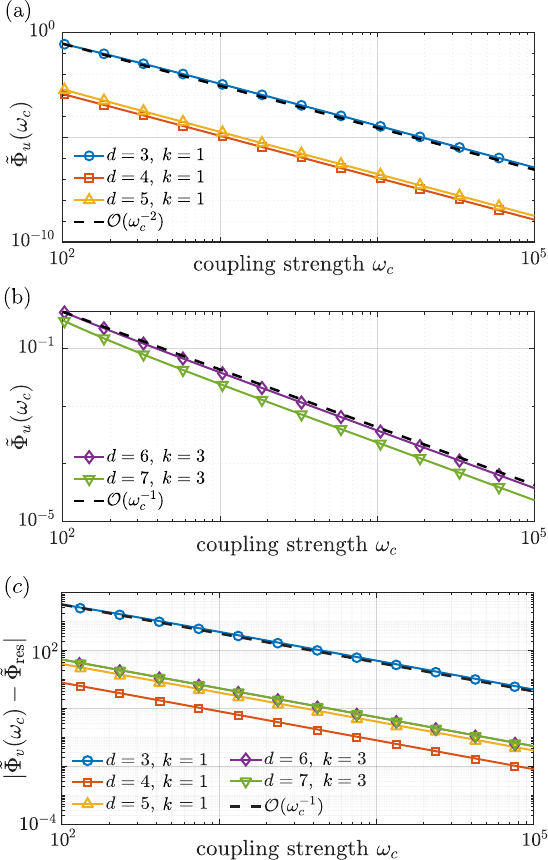}
    \caption{Strong-coupling scaling of the modal EPR contributions.
(a) For a single directly coupled variable, $k=1$, the condition in Eq.~(\ref{eq:fast-detailed-balance}) holds
automatically, and the coupled contribution decays as
$|\tilde{\Phi}_u-0| = \tilde{\Phi}_u =\mathcal{O}(\omega_c^{-2})$.
(b) When Eq.~(\ref{eq:fast-detailed-balance}) is violated, the coupled
contribution exhibits the generic decay
$\tilde{\Phi}_u=\mathcal{O}(\omega_c^{-1})$.
(c) For the same representative systems, the modal EPR approaches the
residual EPR as
$|\tilde{\Phi}_m-\tilde{\Phi}_{\mathrm{res}}|
=\mathcal{O}(\omega_c^{-1})$.
The black dashed lines indicate the predicted asymptotic scalings, while
$d$ and $k$ denote the total number of variables and the number of
directly coupled variables, respectively.}
\label{fig:strong-coupling-scalings}
\end{figure}

When $k=1$, both $\mathbf{A}$ and $\mathbf{D}_u$ are scalars, and
Eq.~(\ref{eq:fast-detailed-balance}) holds automatically. Equivalently,
the antisymmetric block $\mathbf{Q}_{uu}$ vanishes identically.
The coupled contribution therefore reduces to
\begin{equation}
    \tilde{\Phi}_u
    =
    \operatorname{tr}
    \left[
        \mathbf{D}_u^{-1}
        \mathbf{Q}_{uv}
        (\mathbf{S}_m^{-1})_{vv}
        \mathbf{Q}_{uv}^{\top}
    \right].
\end{equation}
From Eqs.~(\ref{eq:Quv-scaling}) and
(\ref{eq:inverse-covariance-scaling}), we have
$\mathbf{Q}_{uv}=\mathcal{O}(\omega_c^{-1})$ and
$(\mathbf{S}_m^{-1})_{vv}
=\mathbf{Z}_0^{-1}+\mathcal{O}(\omega_c^{-1})$. Consequently,
\begin{equation}
    \tilde{\Phi}_u
    =
    \mathcal{O}(\omega_c^{-2}).
    \label{eq:phi-u-k1}
\end{equation}
Thus, for a single directly coupled variable, the coupled EPR contribution
is suppressed one order faster than the generic
$\mathcal{O}(\omega_c^{-1})$ decay, as illustrated in
Fig.~\ref{fig:strong-coupling-scalings}(a).

\subsubsection{Convergence to residual EPR}
\label{app:residual-epr-proof}

We next consider the EPR contribution associated with the residual variables.
Since $\mathbf{Q}_m$ is antisymmetric,
Eq.~(\ref{eq:Quv-scaling}) implies
\begin{equation}
    \mathbf{Q}_{vu}
    =
    -\mathbf{Q}_{uv}^{\top}
    =
    \mathcal{O}(\omega_c^{-1}).
    \label{eq:Qvu-scaling}
\end{equation}
Using the covariance expansion in
Eq.~(\ref{eq:covariance-strong-coupling}), the residual current block
satisfies
\begin{equation}
\begin{split}
    \mathbf{Q}_{vv}
    &=
    \mathbf{C}\mathbf{S}_{uv}
    +
    \mathbf{R}\mathbf{S}_{vv}
    +
    \mathbf{D}_v
    \\
    &=
    \mathbf{R}\mathbf{Z}_0
    +
    \mathbf{D}_v
    +
    \mathcal{O}(\omega_c^{-1}).
\end{split}
\label{eq:Qvv-scaling}
\end{equation}
We therefore define the irreversible-current matrix of the residual
dynamics as
\begin{equation}
    \mathbf{Q}_{\mathrm{res}}
    =
    \mathbf{R}\mathbf{Z}_0+\mathbf{D}_v.
    \label{eq:Q-residual}
\end{equation}
Together with Eq.~(\ref{eq:residual-lyapunov}), this is precisely the
irreversible-current matrix of the Ornstein--Uhlenbeck process with drift
$\mathbf{R}$, diffusion $\mathbf{D}_v$, and covariance $\mathbf{Z}_0$.

Using the modal EPR decomposition derived above,
\begin{equation}
    \tilde{\Phi}_v
    =
    \operatorname{tr}
    \left[
        \mathbf{D}_v^{-1}
        \mathbf{Q}_v
        \mathbf{S}_m^{-1}
        \mathbf{Q}_v^\top
    \right],
\end{equation}
where $\mathbf{Q}_v=(\mathbf{Q}_{vu}\ \mathbf{Q}_{vv})$. Writing the
inverse covariance in block form, the expression inside the trace becomes
\begin{align}
    \mathbf{Q}_v\mathbf{S}_m^{-1}\mathbf{Q}_v^\top
    &=
    \mathbf{Q}_{vu}
    (\mathbf{S}_m^{-1})_{uu}
    \mathbf{Q}_{vu}^{\top}
    \notag\\
    &\quad+
    \mathbf{Q}_{vu}
    (\mathbf{S}_m^{-1})_{uv}
    \mathbf{Q}_{vv}^{\top}
    +
    \mathbf{Q}_{vv}
    (\mathbf{S}_m^{-1})_{vu}
    \mathbf{Q}_{vu}^{\top}
    \notag\\
    &\quad+
    \mathbf{Q}_{vv}
    (\mathbf{S}_m^{-1})_{vv}
    \mathbf{Q}_{vv}^{\top}.
    \label{eq:phi-v-block-expansion}
\end{align}
From Eqs.~(\ref{eq:Qvu-scaling}) and
(\ref{eq:inverse-covariance-scaling}), the first term on the right-hand
side is of order $\omega_c^{-1}$, and both cross terms are also of order
$\omega_c^{-1}$. By contrast, Eqs.~(\ref{eq:Qvv-scaling}) and
(\ref{eq:inverse-covariance-scaling}) give
\begin{equation}
    \mathbf{Q}_{vv}
    (\mathbf{S}_m^{-1})_{vv}
    \mathbf{Q}_{vv}^{\top}
    =
    \mathbf{Q}_{\mathrm{res}}
    \mathbf{Z}_0^{-1}
    \mathbf{Q}_{\mathrm{res}}^{\top}
    +
    \mathcal{O}(\omega_c^{-1}).
\end{equation}
Consequently,
\begin{equation}
\begin{split}
    \tilde{\Phi}_v
    &=
    \operatorname{tr}
    \left[
        \mathbf{D}_v^{-1}
        \mathbf{Q}_{\mathrm{res}}
        \mathbf{Z}_0^{-1}
        \mathbf{Q}_{\mathrm{res}}^{\top}
    \right]
    +
    \mathcal{O}(\omega_c^{-1})
    \\
    &=
    \tilde{\Phi}_{\mathrm{res},m}
    +
    \mathcal{O}(\omega_c^{-1}).
\end{split}
\label{eq:phi-v-limit}
\end{equation}
as depicted in Fig.~\ref{fig:strong-coupling-scalings}(c). Thus,
\begin{equation}
    \tilde{\Phi}_v
    \longrightarrow
    \tilde{\Phi}_{\mathrm{res},m}
    \qquad
    \text{as }
    \omega_c\to\infty.
\end{equation}
 The strong-coupling limit of the residual contribution is therefore the EPR
of the residual dynamics as a whole.

Combining Eqs.~(\ref{eq:phi-u-limit}) and
(\ref{eq:phi-v-limit}), the contribution of every nonzero coupling mode
satisfies
\begin{equation}
\begin{split}
    \tilde{\Phi}_m(\omega_c)
    &=
    \tilde{\Phi}_u(\omega_c)
    +
    \tilde{\Phi}_v(\omega_c)
    \\
    &=
    \tilde{\Phi}_{\mathrm{res},m}
    +
    \mathcal{O}(\omega_c^{-1}).
\end{split}
\label{eq:modal-residual-limit}
\end{equation}
Hence,
\begin{equation}
    \lim_{\omega_c\to\infty}
    \tilde{\Phi}_m(\omega_c)
    =
    \tilde{\Phi}_{\mathrm{res},m}
\end{equation}
for every nonzero mode. Strong coupling therefore suppresses the EPR
associated with the directly coupled variables, while the residual
variables may continue to sustain a finite irreversible current.

The strong-coupling limit can also be viewed in the context of
time-scale separation and coarse-graining in stochastic thermodynamics,
where strongly damped variables are eliminated to obtain an effective
dynamics for the remaining variables~\cite{Wang2016,Busiello2020}. Here, the same separation
arises through increasing coupling strength, while the analysis additionally
determines the leading finite-coupling correction to the EPR.

Let $n_0$ denote the number of zero modes, and label them by
$m=1,\ldots,n_0$. These modes are unaffected by the coupling, whereas the
remaining modes converge to their residual contributions. The total EPR
therefore satisfies
\begin{equation}
    \Phi_n(\omega_c)
    =
    \sum_{m=1}^{n_0}
    \tilde{\Phi}_m(0)
    +
    \sum_{m=n_0+1}^{n}
    \tilde{\Phi}_{\mathrm{res},m}
    +
    \mathcal{O}(\omega_c^{-1}),
    \label{eq:total-residual-limit}
\end{equation}
and consequently
\begin{equation}
    \lim_{\omega_c\to\infty}\Phi_n(\omega_c)
    =
    \sum_{m=1}^{n_0}
    \tilde{\Phi}_m(0)
    +
    \sum_{m=n_0+1}^{n}
    \tilde{\Phi}_{\mathrm{res},m}.
\end{equation}

For a connected coupling topology, there is a single zero mode whose
contribution is equal to the EPR $\Phi_1$ of one uncoupled subsystem. In
this case,
\begin{equation}
    \lim_{\omega_c\to\infty}\Phi_n(\omega_c)
    =
    \Phi_1
    +
    \sum_{m=2}^{n}
    \tilde{\Phi}_{\mathrm{res},m}.
\end{equation}
Thus, the strong-coupling limit consists of the unchanged zero-mode
contribution together with the residual EPR of each nonzero mode.

\subsection{First-order coefficients}
\label{app:first-order-coefficient}

We now determine the leading correction governing the approach of each
nonzero mode to its residual EPR. As before, we suppress the mode index.
The first-order coefficient contains two non-negative contributions
associated with the suppressed variables and a remaining contribution from
the change in the residual current.

To obtain the first-order term in $\tilde{\Phi}_u$, we refine the expansion
of the coupled covariance block in
Eq.~(\ref{eq:covariance-strong-coupling}) as
\begin{equation}
    \mathbf{S}_{uu}
    =
    \frac{\mathbf{X}_1}{\omega_c}
    +
    \frac{\mathbf{X}_2}{\omega_c^2}
    +
    \mathcal{O}(\omega_c^{-3}).
    \label{eq:Suu-second-order}
\end{equation}
Using Eq.~(\ref{eq:A0X1-cancellation}), the corresponding current block is
\begin{equation}
    \mathbf{Q}_{uu}
    =
    \frac{\mathbf{U}_1}{\omega_c}
    +
    \mathcal{O}(\omega_c^{-2}),
    \label{eq:Quu-first-order}
\end{equation}
where
\begin{equation}
       \mathbf{U}_1
    \equiv
    \mathbf{A}_0\mathbf{X}_2
    +
    \mathbf{A}\mathbf{X}_1
    +
    \mathbf{B}\mathbf{Y}_1^\top.
\end{equation}
Moreover,
$(\mathbf{S}_m^{-1})_{uu}
=\omega_c\mathbf{X}_1^{-1}+\mathcal{O}(1)$.
All terms in $\tilde{\Phi}_u$ involving $\mathbf{Q}_{uv}$ are of order
$\omega_c^{-2}$. This gives the expansion in Eq.~(\ref{phi_u}), with
\begin{equation}
    K_u
    =
    \operatorname{tr}
    \left[
        \mathbf{D}_u^{-1}
        \mathbf{U}_1
        \mathbf{X}_1^{-1}
        \mathbf{U}_1^\top
    \right]
    \geq0.
    \label{eq:Ku}
\end{equation}
For a single directly coupled variable, $\mathbf{Q}_{uu}$ vanishes
identically and hence $K_u=0$, consistent with the faster decay derived
above.

\subsubsection{Current overlap}
\label{app:current-alignment}

We next expand the residual contribution $\tilde{\Phi}_v$ and identify
the origin of its first-order terms. Completing the square in
Eq.~(\ref{eq:phi-v-block-expansion}) gives
\begin{equation}
\begin{split}
    \tilde{\Phi}_v
    &=
    \operatorname{tr}
    \left[
        \mathbf{D}_v^{-1}
        \widehat{\mathbf{Q}}_{vu}
        (\mathbf{S}_m^{-1})_{uu}
        \widehat{\mathbf{Q}}_{vu}^{\top}
    \right]
    \\
    &\quad+
    \operatorname{tr}
    \left[
        \mathbf{D}_v^{-1}
        \mathbf{Q}_{vv}
        \mathbf{S}_{vv}^{-1}
        \mathbf{Q}_{vv}^{\top}
    \right],
\end{split}
\label{eq:phi-v-completed-square}
\end{equation}
where
\begin{equation}
    \widehat{\mathbf{Q}}_{vu}
    =
    \mathbf{Q}_{vu}
    +
    \mathbf{Q}_{vv}
    (\mathbf{S}_m^{-1})_{vu}
    \left[(\mathbf{S}_m^{-1})_{uu}\right]^{-1}.
\end{equation}
Here we used the block-inversion identity
\begin{equation}
    (\mathbf{S}_m^{-1})_{vv}
    -
    (\mathbf{S}_m^{-1})_{vu}
    \left[(\mathbf{S}_m^{-1})_{uu}\right]^{-1}
    (\mathbf{S}_m^{-1})_{uv}
    =
    \mathbf{S}_{vv}^{-1}.
\end{equation}

From the strong-coupling expansion of the covariance,
\begin{align}
    (\mathbf{S}_m^{-1})_{uu}
    &=
    \omega_c\mathbf{X}_1^{-1}
    +
    \mathcal{O}(1),
    \\
    (\mathbf{S}_m^{-1})_{vu}
    \left[(\mathbf{S}_m^{-1})_{uu}\right]^{-1}
    &=
    -\frac{1}{\omega_c}
    \mathbf{Z}_0^{-1}\mathbf{Y}_1^\top
    +
    \mathcal{O}(\omega_c^{-2}).
\end{align}
Furthermore,
\begin{align}
    \mathbf{Q}_{vu}
    &=
    \frac{
        \mathbf{C}\mathbf{X}_1
        +
        \mathbf{R}\mathbf{Y}_1^\top
    }{\omega_c}
    +
    \mathcal{O}(\omega_c^{-2}),
    \\
    \mathbf{Q}_{vv}
    &=
    \mathbf{Q}_{\mathrm{res}}
    +
    \mathcal{O}(\omega_c^{-1}).
\end{align}
Using
$\mathbf{Q}_{\mathrm{res}}\mathbf{Z}_0^{-1}
=\mathbf{R}+\mathbf{D}_v\mathbf{Z}_0^{-1}$, we obtain
\begin{equation}
    \widehat{\mathbf{Q}}_{vu}
    =
    \frac{\mathbf{H}_1}{\omega_c}
    +
    \mathcal{O}(\omega_c^{-2}),
\end{equation}
where
\begin{equation}
    \mathbf{H}_1
    =
    \mathbf{C}\mathbf{X}_1
    -
    \mathbf{D}_v\mathbf{Z}_0^{-1}\mathbf{Y}_1^\top.
\end{equation}
Hence, the first term in Eq.~(\ref{eq:phi-v-completed-square}) satisfies
\begin{equation}
    \operatorname{tr}
    \left[
        \mathbf{D}_v^{-1}
        \widehat{\mathbf{Q}}_{vu}
        (\mathbf{S}_m^{-1})_{uu}
        \widehat{\mathbf{Q}}_{vu}^{\top}
    \right]
    =
    \frac{K_v}{\omega_c}
    +
    \mathcal{O}(\omega_c^{-2}),
\end{equation}
where
\begin{equation}
    K_v
    =
    \operatorname{tr}
    \left[
        \mathbf{D}_v^{-1}
        \mathbf{H}_1
        \mathbf{X}_1^{-1}
        \mathbf{H}_1^\top
    \right]
    \geq 0.
    \label{eq:Kv}
\end{equation}

To interpret the remaining term, we define the normalized irreversible
current within the residual variables as
\begin{equation}
    \mathbf{J}_{vv}(\omega_c)
    =
    \mathbf{D}_v^{-1/2}
    \mathbf{Q}_{vv}
    \mathbf{S}_{vv}^{-1/2}.
    \label{eq:normalized-residual-current}
\end{equation}
Using the strong-coupling expansion in Eq.~(\ref{eq:J_expansion}),
\begin{equation}
    \mathbf{J}_{\mathrm{res}}
    =
    \mathbf{D}_v^{-1/2}
    \mathbf{Q}_{\mathrm{res}}
    \mathbf{Z}_0^{-1/2},
    \qquad
    \tilde{\Phi}_{\mathrm{res}}
    =
    \|\mathbf{J}_{\mathrm{res}}\|_F^2.
\end{equation}
Therefore,
\begin{equation}
\begin{split}
    \operatorname{tr}
    \left[
        \mathbf{D}_v^{-1}
        \mathbf{Q}_{vv}
        \mathbf{S}_{vv}^{-1}
        \mathbf{Q}_{vv}^{\top}
    \right]
    &=
    \|\mathbf{J}_{vv}(\omega_c)\|_F^2
    \\
    &=
    \tilde{\Phi}_{\mathrm{res}}
    +
    \frac{2}{\omega_c}
    \left\langle
        \mathbf{J}_{\mathrm{res}},
        \mathbf{J}_1
    \right\rangle_F
    +
    \mathcal{O}(\omega_c^{-2}).
\end{split}
\label{eq:residual-current-alignment}
\end{equation}

Combining the two terms gives the residual expansion in
Eq.~(\ref{phi_v}), with the sign-indeterminate contribution $\tilde K$
given by Eq.~(\ref{eq:inner-product}).

It is useful to note that evaluating the coefficients $K_u, K_v, \tilde{K}$ does
not require solving the full $d$-dimensional Lyapunov equation for
$\mathbf{S}_m(\omega_c)$. Instead, the required covariance coefficients are
obtained successively from lower-dimensional matrix equations for the
coupled and residual blocks. In particular, $\mathbf{X}_1$ and
$\mathbf{X}_2$ are determined within the $k$-dimensional coupled subspace,
while $\mathbf{Z}_0$ and $\mathbf{Z}_1$ follow from Lyapunov equations in
the $(d-k)$-dimensional residual subspace, and $\mathbf{Y}_1$ follows
directly from Eq.~(\ref{eq:leading-covariance-coefficients}). Thus, the
strong-coupling response can be determined without constructing the full
finite-coupling covariance, which can be advantageous when the coupled or
residual subspaces are substantially smaller than the full system.

\subsection{Example of asymptotic EPR increase}
\label{app:epr-increase}

\begin{figure}
    \centering
    \includegraphics[width=\linewidth]{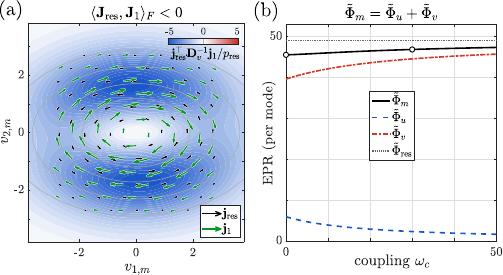}
    \caption{
    Example of asymptotic EPR increase in a representative higher-dimensional
    mode.
    (a) Residual current $\mathbf j_{\rm res}$ and leading finite-coupling
    correction $\mathbf j_1$, shown over their signed overlap density.
    The negative overlap indicates that the finite-coupling correction opposes
    the residual circulation.
    (b) Modal EPR $\tilde{\Phi}_m$ as a function of $\omega_c$.
    For this parameter choice, $\tilde{\Phi}_m$ approaches
    $\tilde{\Phi}_{\rm res}$ from below and therefore increases in the
    strong-coupling regime, with
    $\tilde{\Phi}_{\rm res}>\tilde{\Phi}_m(0)$.
    \textit{Simulation parameters:}
    $a=20$, $p=7.5$, $r_1=0.25$, $r_2=0.75$, $q_1=5$, $q_2=2$,
    $b_1=b_2=1.67$ and $c_1=c_2=-1.67$.
    }
    \label{fig:appendix_current}
\end{figure}

When $\tilde K$ is sufficiently negative, the EPR increases
asymptotically. To illustrate this explicitly, we return to the
four-dimensional model introduced in
Eq.~(\ref{eq:representative_example}) and choose parameters for which the
finite-coupling correction opposes the residual current. In this case,
$\langle \mathbf J_{\rm res},\mathbf J_1\rangle_F<0$ and
$\tilde K<-(K_u+K_v)$, and hence $K<0$. Consequently, the modal EPR
approaches the residual EPR from below and increases with $\omega_c$ in the
strong-coupling regime, as shown in Fig.~\ref{fig:appendix_current}.
For this parameter choice, $\tilde{\Phi}_{\rm res}=49$ exceeds the
uncoupled value $\tilde{\Phi}_m(0)\simeq45.51$, so sufficiently strong
coupling also leads to greater dissipation than in the uncoupled system.

\section{Different effective temperatures}
\label{app:effective-temperatures}

The convergence to the residual EPR derived in
Appendix~\ref{app:residual-convergence} relies on the cancellation in
Eq.~(\ref{eq:A0X1-cancellation}). Within the present setup,
$\mathbf A_0$ is diagonal in the coupling eigenbasis by construction.
The cancellation therefore holds when the diffusion block associated with
the nonzero coupling eigendirections is also diagonal in this basis. For
equal effective temperatures of the directly coupled variables,
transformation to the coupling eigenbasis preserves this diagonal diffusion
structure. The different coupling eigendirections are then statistically
independent, and for every nonzero coupling direction the leading covariance
coefficient satisfies Eq.~(\ref{eq:X1-lyapunov}), which reduces to
Eq.~(\ref{eq:A0X1-cancellation}). Consequently, the leading contribution
to $\mathbf Q_{uu}$ vanishes, giving the scaling in
Eq.~(\ref{eq:Quu-scaling}) and ultimately
$\tilde{\Phi}_u\to0$ as in Eq.~(\ref{eq:phi-u-limit}).

When the directly coupled variables of different subsystems are maintained
at different effective temperatures, transformation to the coupling
eigenbasis generally makes the diffusion matrix associated with these
variables non-diagonal. If several nonzero coupling modes are present,
the resulting off-diagonal terms can occur within $\mathbf D_u$ itself.
When this destroys the diagonal structure shared by $\mathbf A_0$ and
$\mathbf D_u$ required for Eq.~(\ref{eq:A0X1-cancellation}),
$\mathbf Q_{uu}$ can retain $\mathcal O(1)$ components in the
strong-coupling limit. Since the covariance associated with the nonzero
coupling modes scales as $\mathcal O(\omega_c^{-1})$, the corresponding
components of the inverse covariance scale as $\mathcal O(\omega_c)$.
The coupled-variable contribution to the EPR can therefore contain terms
of the form
\begin{equation}
    \operatorname{tr}\!\left[
        \mathbf D_u^{-1}
        \mathbf Q_{uu}
        (\mathbf S^{-1})_{uu}
        \mathbf Q_{uu}^{\top}
    \right]
    =
    \mathcal O(\omega_c),
    \label{eq:unequal-temp-Quu}
\end{equation}
and the total EPR need not converge to the residual EPR.

Off-diagonal diffusion terms can also couple a nonzero coupling mode to a
zero-eigenvalue collective mode. In this case, residual-EPR convergence
can fail through mixed irreversible-current contributions rather than
through $\mathbf Q_{uu}$. By contrast, off-diagonal diffusion confined
entirely to the residual variables modifies the residual covariance and
EPR but does not invalidate convergence to the residual EPR. Related
persistence of dissipation in strongly constrained fast variables has been
identified in underdamped Brownian systems coupled to reservoirs at
different temperatures \citep{ChunNoh2015}.

For the two-system example discussed in the main text, the diffusion matrix
of the directly coupled variables $(y_1,y_2)^\top$ is
$\operatorname{diag}(D_y^{(1)},D_y^{(2)})$. In the eigenbasis of the
coupling matrix, this diffusion matrix is non-diagonal whenever
$D_y^{(1)}\neq D_y^{(2)}$. Here there is only one nonzero coupling mode,
corresponding to the relative coordinate, together with an unaffected
collective mode. The corresponding $\mathbf D_u$ block is therefore
one-dimensional, and $\mathbf Q_{uu}$ vanishes identically, so the
contribution in Eq.~(\ref{eq:unequal-temp-Quu}) is absent. Instead, the
off-diagonal diffusion term couples the relative and collective modes,
allowing the corresponding mixed irreversible current to remain finite in
the strong-coupling limit. Since the inverse covariance along the relative
mode grows as $\mathcal O(\omega_c)$, this mixed current produces the
linear strong-coupling contribution in
Eq.~(\ref{eq:temperature-EPR}).

When the coupled variables have the same effective temperature, the
off-diagonal diffusion term vanishes in the coupling eigenbasis and the
leading contribution is removed, recovering convergence to the residual
EPR. For different effective temperatures, the surviving linear term
corresponds physically to the coupling-mediated heat flux between the two
noise environments.

\nocite{*}

\bibliography{newbib}
\end{document}